\documentclass{aa}

\usepackage{graphicx}
\usepackage{txfonts}
\usepackage{hyperref}
\usepackage[noabbrev]{cleveref}
\begin{document} 

   \title{Focal-plane wavefront sensing with the vector Apodizing Phase Plate}
   
   \author{S.P. Bos\inst{1}, D.S. Doelman\inst{1}, J. Lozi\inst{2}, O. Guyon\inst{2,3,4,5}, C.U. Keller\inst{1},  K.L. Miller\inst{1},\\ N. Jovanovic\inst{6}, F. Martinache\inst{7}, F. Snik\inst{1}}
   \institute{Leiden Observatory, Leiden University, P.O. Box 9513, 2300 RA Leiden, The Netherlands\\
              	\email{stevenbos@strw.leidenuniv.nl}
		\and	
		National Astronomical Observatory of Japan, Subaru Telescope, National Institute of Natural Sciences, Hilo, HI 96720, USA \\ 
	         \and
        		 Steward Observatory, University of Arizona, 933 N. Cherry Ave, Tucson, AZ 85721, USA\\
         	\and
             	College of Optical Sciences, University of Arizona, 1630 E. University Blvd., Tucson, AZ 85721, USA \\
		\and
		Astrobiology Center, National Institutes of Natural Sciences, 2-21-1 Osawa, Mitaka, Tokyo, Japan \\  
		\and 
		Department of Astronomy, California Institute of Technology, 1200 E. California Blvd., Pasadena, CA 91125, USA\\   
		\and
		Observatoire de la Cote d'Azur, Boulevard de l'Observatoire, Nice, 06304, France     
             }

   \date{Received June 12, 2019; accepted September 16, 2019}

 
  \abstract
   {One of the key limitations of the direct imaging of exoplanets at small angular separations are quasi-static speckles that originate from evolving non-common path aberrations (NCPA) in the optical train downstream of the instrument's main wavefront sensor split-off.}
   {In this article we show that the vector-Apodizing Phase Plate (vAPP) coronagraph can be designed such that the coronagraphic point spread functions (PSFs) can act as a wavefront sensor to measure and correct the (quasi-)static aberrations, without dedicated wavefront sensing holograms nor modulation by the deformable mirror. The absolute wavefront retrieval is performed with a non-linear algorithm.}
   {The focal-plane wavefront sensing (FPWFS) performance of the vAPP and the algorithm are evaluated with numerical simulations, to test various photon and read noise levels, the sensitivity to the 100 lowest Zernike modes and the maximum wavefront error (WFE) that can be accurately estimated in one iteration. We apply these methods to the vAPP within SCExAO, first with the internal source and subsequently on-sky.}
   {In idealised simulations we show that for $10^7$ photons the root-mean-square (RMS) WFE can be reduced to $\sim\lambda/1000$, which is 1 nm RMS in the context of the SCExAO system. We find that the maximum WFE that can be corrected in one iteration is $\sim\lambda/8$ RMS or $\sim$200 nm RMS (SCExAO). Furthermore, we demonstrate the SCExAO vAPP capabilities by measuring and controlling the lowest 30 Zernike modes with the internal source and on-sky. On-sky, we report a raw contrast improvement of a factor $\sim$2 between 2 and 4 $\lambda/D$ after 5 iterations of closed-loop correction. When artificially introducing 150 nm RMS WFE, the algorithm corrects it within 5 iterations of closed-loop operation.}
   {FPWFS with the vAPP's coronagraphic PSFs is a powerful technique since it integrates coronagraphy and wavefront sensing, eliminating the need for additional probes and thus resulting in a $100\%$ science duty cycle and maximum throughput for the target.}


\authorrunning{S. P. Bos et al.}
\maketitle

\section{Introduction}\label{sec:introduction}
The exploration of circumstellar environments at small angular separations by means of direct imaging is crucial for the detection and characterization of exoplanets. The challenges that need to be overcome are that of high contrast and small angular separation: Earth around the Sun at 10 pc will have an angular separation of $\sim$100 mas and a contrast in the visible ($\sim$0.3-1 $\mu$m) of $\sim$$10^{-10}$ \citep{traub2010direct}. \\
\indent The current generation of ground-based high-contrast imaging instruments SPHERE \citep{beuzit2019sphere}, GPI \citep{macintosh2014first}, SCExAO \citep{jovanovic2015subaru} and the upcoming MagAO-X (\citealt{males2018magao}; \citealt{close2018optical}) are pushing towards contrasts of $\sim$$10^{-6}$ at angular separations of 200 mas after post-processing in the near-infrared (0.95-2.3 $\mu$m; \citealt{vigan2015high}). These instruments are equipped with extreme adaptive optics systems to flatten the wavefront after the turbulent atmosphere, coronagraphs to suppress the star light and contrast-enhancing post-processing techniques that employ some form of diversity such as angular differential imaging \citep{marois2006angular}, reference star differential imaging \citep{ruane2019reference}, spectral differential imaging \citep{sparks2002imaging} and polarimetric differential imaging \citep{snik2013astronomical}. The latter two techniques can also be used as a characterization diagnostic. Medium- and high-resolution integral-field spectroscopy can be used to detect atomic and molecular lines from a planet's atmosphere (e.g. \citealt{haffert2019two} and \citealt{hoeijmakers2018medium}) while polarimetry can be used to detect cloud structures (\citealt{stam2004using}; \citealt{de2011characterizing}; \citealt{van2017combining}).\\

\begin{figure}
\centering
\includegraphics[width=\hsize]{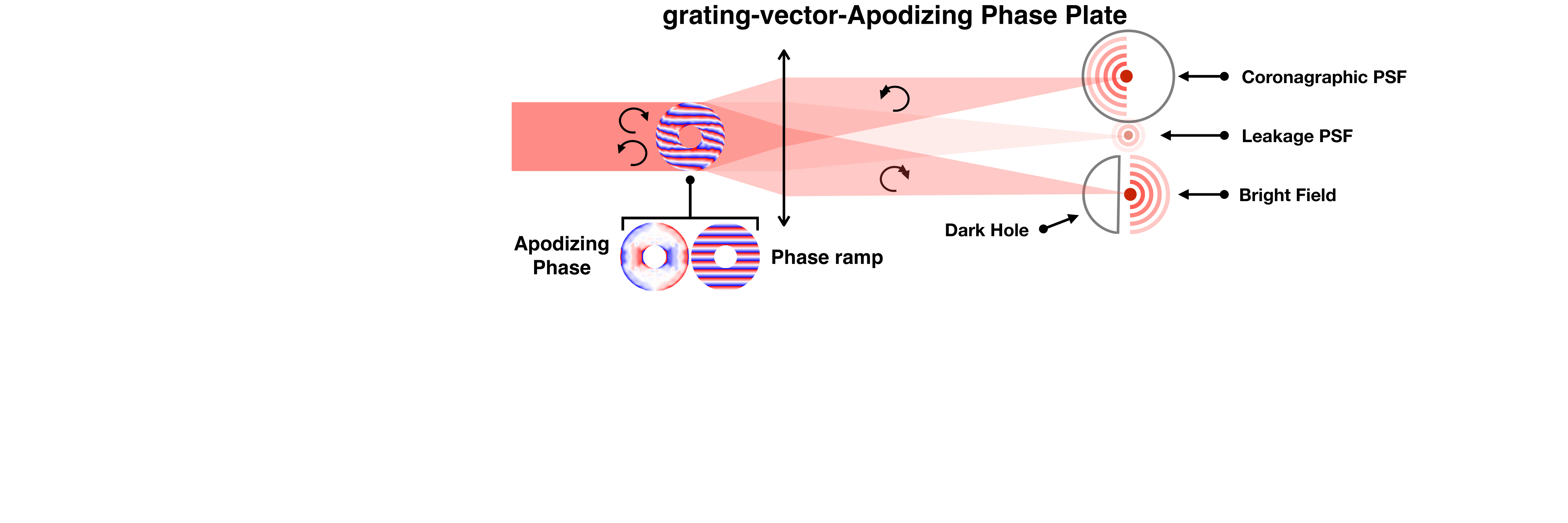}
\caption{Working principle of the grating-vector-Apodizing Phase Plate. The vAPP is a half-wave retarder pupil-plane optic with a spatially varying fast-axis orientation. The varying fast-axis orientation induces the phase through the geometric phase on the circular polarization states. These polarization states receive the opposite phase and therefore have flipped coronagraphic point spread functions (PSFs). The PSFs are spatially separated by adding a phase ramp to the design. Any offsets from half-wave retardance within the optic reduces the efficiency with which the light will be transferred to the coronagraphic PSFs and results in a leakage PSF. }
\label{fig:gvAPP_example}
\end{figure}

\indent The coronagraph relevant for the present work is the vector-Apodizing Phase Plate (vAPP; \citealt{snik2012vector}; \citealt{otten2017sky}), which manipulates the phase in the pupil-plane such that in selected regions in the focal-plane the starlight is cancelled; these areas are referred to as the dark holes. The phase is induced through the achromatic geometric phase (\citealt{pancharatnam1956generalized}; \citealt{berry1987adiabatic}) on the circular polarization states by a half-wave liquid-crystal layer with a varying fast-axis orientation. The two circular polarization states both receive equal, but opposite phases, resulting in two coronagraphic PSFs with opposite dark holes as shown in \autoref{fig:gvAPP_example}. The geometric phase is inherently achromatic as it depends on geometric effects, but the efficiency with which the light acquires the phase is determined by the retardance of the liquid-crystal layer. Retardance offsets from half-wave result in leakage; light that has not acquired the desired phase will form a non-coronagraphic PSF based on the aperture geometry. Generally, vAPP coronagraphs are designed to have minimal leakage over a broad wavelength range. High leakage will affect coronagraphic performance as light from the leaked PSF will contaminate the dark hole. In the simplest and most common implementation the two coronagraphic PSFs are spatially separated with a polarization-sensitive grating \citep{oh2008achromatic} that is integrated in the phase design. These coronagraphs are known as grating-vAPPs \citep{otten2014vector} and are mainly used for operation with narrowband filters or integral-field spectrographs to prevent smearing by the grating. In this article the grating-vAPP will be referred to as a vAPP. The vAPP has been put on-sky with several instruments: CHARIS/SCExAO \citep{doelman2017patterned}, MagAO/Clio2 \citep{otten2017sky}, LMIRCAM/LBT \citep{doelman2017patterned} and LEXI \citep{haffert2018sky}. Furthermore, vAPPs have been designed for the following instruments: HiCIBaS \citep{cote2018precursor},  MagAO-X \citep{miller2018focal}, ERIS \citep{boehle2018cryogenic}, METIS \citep{kenworthy2018review} and MICADO \citep{davies2018micado}. \\

One of the key limitations of current high contrast imaging instruments that limit them to contrasts above $\sim10^{-6}$ within 300 mas are quasi-static speckles that originate from slowly-evolving instrumental aberrations caused by changing temperature, humidity and gravity vector during observations (\citealt{martinez2012speckle}; \citealt{martinez2013speckle}; \citealt{n2016zelda}; \citealt{vigan2018sky}). When these aberrations occur in the optical train downstream of the main wavefront sensor split-off, they will not be sensed and therefore can not be corrected. Additional focal-plane wavefront sensing (FPWFS) with the science detector is a highly desirable solution to these non-common path aberrations (NCPA). Besides eliminating the NCPAs, a FPWS can also address chromatic errors between the main sensing and science channels. Another advantage of FPWFS is that it has been shown by \cite{guyon2005limits} to be able to reach high sensitivities for all spatial frequencies, only being bested in sensitivity by the Zernike wavefront sensor (\citealt{n2013calibration}; \citealt{doelman2019simultaneous}). A notable FPWFS is the Self-Coherent Camera (SCC; \citealt{baudoz2005self}; \citealt{galicher2008wavefront}; \citealt{mazoyer2013estimation}). This is a WFS that is combined with a coronagraph, which uses focal-plane optics to block starlight. It operates by placing a small hole in the pupil-plane Lyot stop outside of the geometric pupil where the scattered starlight is located. This hole creates a reference beam that will generate high spatial frequency fringes in the focal-plane image, which can be used to determine the full electric field. The SCC has a 100$\%$ science duty cycle, but requires a high focal-plane sampling to resolve the fringes, and optics of a sufficient size to accommodate the reference beam. FPWS has also been conducted by using vAPPs. Previous work focused on adding additional holograms in the focal-plane that encode wavefront information \citep{wilby2017coronagraphic} or directly probe the electric field \citep{por2016focal}; these will not be considered here.\\

\indent An overview of FPWFS techniques can be found in \cite{Jovanovic2018}. Three FPWFS and control methods particularly relevant are: 
\begin{itemize}
\item The COronagraphic Focal-plane waveFront Estimation for Exoplanet detection (COFFEE; \citealt{sauvage2012coronagraphic}; \citealt{paul2013high}) wavefront sensor is an extension of classical phase diversity (\citealt{gonsalves1982phase};  \citealt{paxman1992joint}) to coronagraphic imaging. Aberrations in a physical model of the coronagraphic system are fitted to two focal-plane images, one of which has a known phase diversity (e.g. defocus). The method has been demonstrated in the lab \citep{paul2013coronagraphic} and on the SPHERE system using the internal calibration source \citep{paul2014compensation}. Recent extensions enable COFFEE to measure phase in long-exposure images affected by residual turbulence \citep{herscovici2017analytic}, and measure both phase and amplitude \citep{herscovici2018experimental}.

\item An interferometric approach to FPWFS is the the Asymmetric Pupil Fourier Wavefront Sensor (APF-WFS; \citealt{martinache2013asymmetric}). The APF-WFS assumes the small aberration regime to use a Fourier analysis of focal-plane images to determine pupil-plane phase aberrations. The image is formed by an asymmetric pupil, which enables the full phase determination. The theory behind this technique will be more extensively discussed in \autoref{sec:theory}. The wavefront sensor has been demonstrated on-sky in closed-loop operation controlling the lowest-order Zernike modes \citep{martinache2016closed}, also in the context of controlling the "island effect" \citep{n2018calibration}.  

\item The third method is linear dark-field control (LDFC; \citealt{guyon2017spectral}) and more specifically the spatial LDFC variant \citep{miller2017spatial}. The idea behind spatial LDFC is to measure and control small aberrations that pollute the dark hole (created by the vAPP or other techniques) by measuring the response of the bright field (see \autoref{fig:gvAPP_example}) relative to a reference state. This response is approximately linear for small aberrations. In \cite{miller2017spatial} it was shown to work for one-sided dark holes and modes that have a response in the bright field and the dark hole. But, in this case, there also is a spatial null-space consisting of the modes that do not have a response in the bright field but do pollute the dark hole. The authors overcame this problem by using the vAPP \citep{miller2018focal}, where the bright field of one coronagraphic PSF covers the dark hole of the other. This has been shown to work in the laboratory \citep{miller2018development}, but still has a significant null space for most vAPP designs - it is insensitive to the even pupil phase modes - as will be shown and addressed in \autoref{sec:theory}.
\end{itemize}
COFFEE and the APF-WFS both suffer from low duty cycles as the science observations have to be stopped for the phase diversity probes or moving the asymmetric mask in and out the beam. LDFC, on the other hand, has a $100\%$ duty cycle, but has currently only been considered for vAPPs with a significant null space - the even pupil phase modes - only works in the small aberration regime and does not perform an absolute phase measurement. In \autoref{sec:theory} we combine the APF-WFS and LDFC with vAPPs, eliminating the null space and improving the duty cycle to $100\%$. In \autoref{sec:model} we present a non-linear algorithm similar to COFFEE that can perform absolute phase retrieval. In \autoref{sec:simulations} we explore the theoretical FPWFS performance of the vAPP and the non-linear algorithm with simulations for the vAPP installed at SCExAO. In \autoref{sec:demonstration} we demonstrate the method first with the internal source and subsequently on sky. In \autoref{sec:conclusions} we discuss the results and present the conclusions. 


\section{Theory}\label{sec:theory} 
\subsection{Phase retrieval}
\begin{figure*}
\centering
\includegraphics[width=17cm]{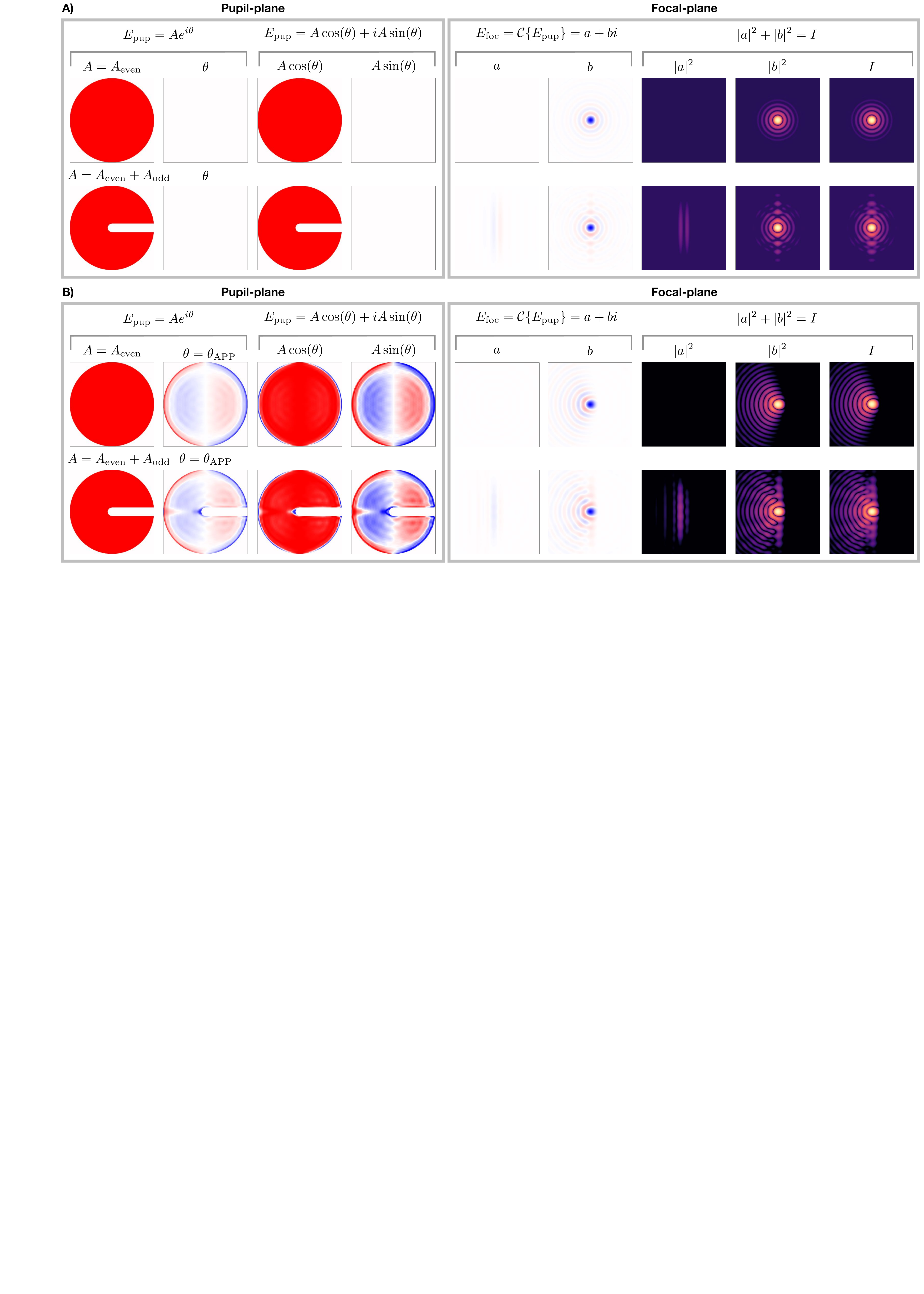}
\caption{Focal- and pupil-plane quantities for (a) a symmetric and asymmetric pupil, and (b) vAPPs designed for these pupils. The columns in the pupil-plane box show (from left to right) the amplitude, phase, real and imaginary electric field. In the focal-plane box, the columns show the real and imaginary electric fields, the power in the real and imaginary electric fields, and the total power.}
\label{fig:theory_fig_B}
\end{figure*}

\begin{figure}
\centering
\includegraphics[width=\hsize]{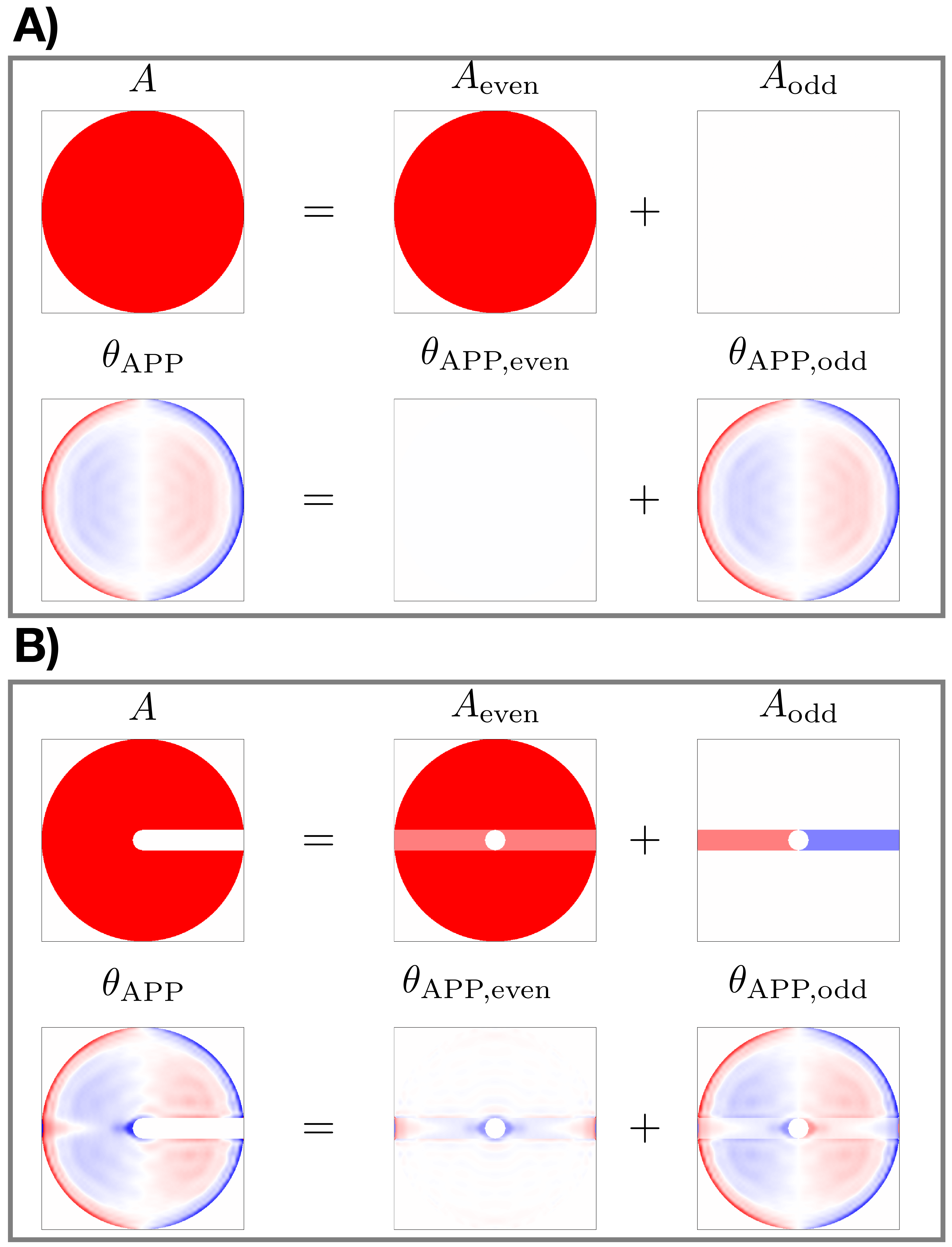}
\caption{Decomposition of the pupil-plane phase and amplitude into their even and odd constituents for the two APP designs in \autoref{fig:theory_fig_B} b. (a) An APP designed for a symmetric aperture. (b) An APP designed for an asymmetric aperture.}
\label{fig:APP_symmetries}
\end{figure}

Phase retrieval techniques in astronomy are concerned with sensing pupil-plane phase aberrations by analysing focal-plane images. These techniques require a unique response in the focal-plane intensity for every phase mode and sign of its modal coefficient (the amount of wavefront error in the specific phase mode). For symmetric pupils such a unique response does not exist for the sign of the modal coefficients of even phase modes. For example, it is easy to determine how an optic needs to be moved to correct for tip/tilt (odd Zernike mode). But when the Point Spread Function (PSF) is defocussed (even Zernike mode), it is not immediately clear in which way the optic needs to be moved to bring the PSF back into focus. We will demonstrate the origin of this well-known sign ambiguity (\citealt{gonsalves1982phase};  \citealt{paxman1992joint}) in this section. \\
\indent The electric field in the pupil-plane consists of an amplitude component $A(r)$ and a phase component $\theta(r)$:
\begin{align}\label{eq:E_pup_tot}
E_{\text{pup}}(r) &= A(r) e^{i \theta(r)} \\ 
			 &= A(r)\cos[\theta(r)] + i A\sin[\theta(r)]. \label{eq:pup_real_imag} 
\end{align}
Here, the electric field $E_{\text{pup}}$ and its constituents, e.g. ${A}$ and ${\theta}$, are 2D entities where the position vector ${r}$, defined from the center of the pupil, is omitted from here on. The focal-plane electric field ${E}_{\text{foc}}(x)$ is formed by propagating ${E}_{\text{pup}}$ using the Fraunhofer propagation operator $\mathcal{C} \{ \cdot \} \propto \frac{1}{i} \mathcal{F}\{ \cdot \} $ \citep{goodman2005introduction}. We use the Fraunhofer propagator, instead of just the Fourier transform, because it is the physically correct propagator. 
\begin{align} 
{E}_{\text{foc}}(x) &= \mathcal{C}\{ {E}_{\text{pup}}\} \\
	       &=  \mathcal{C}\{{A}\cos(\theta)\} + \mathcal{C}\{i {A} \sin(\theta) \} \\
	       &= {a}(x) + i {b}(x),
\end{align}
here ${a}(x)$ and ${b}(x)$ are the real and imaginary components of ${E}_{\text{foc}}(x)$, respectively, and generally contain a mixture of $\mathcal{C}\{{A}\cos(\theta)\}$ and $\mathcal{C}\{i {A} \sin(\theta) \}$. The focal-plane coordinates are denoted by x and are omitted from here on as well. The focal-plane intensity ${I}_{\text{foc}}$ or PSF is subsequently given by:
\begin{align}\label{eq:intensity}
{I}_{\text{foc}} &= |{E}_{\text{foc}}|^2 \\
		             &= |{a}|^2 + |{b}|^2\label{eq:intensity_real_imag}
\end{align}
\begin{table*}
\caption{Fraunhofer propagation symmetry properties \citep{goodman2005introduction}, these are the Fourier properties multiplied with a factor $-i$ (${E}_{\text{foc}} = \mathcal{C} \{ {E}_{\text{pup}}\} \propto \frac{1}{i} \mathcal{F}\{ {E}_{\text{pup}}\}$).}
\vspace{2.5mm}
\centering
\begin{tabular}{ccc|cc}
\hline
\hline
Pupil-plane electric field &  & & Focal-plane electric field &  \\ 
${E}_{\text{pup}} = {A} \cos({\theta}) + i {A} \sin({\theta}) $ & & & ${E}_{\text{foc}} = {a} + i {b}$ &  \\ \hline
Term & Term symmetry & ${A}$, $\theta$ symmetry & Term & Term symmetry \\ \hline
${A} \cos({\theta})$ & Even & $({A}_{\text{even}}, {\theta}_{\text{even}})$, $({A}_{\text{even}}, {\theta}_{\text{odd}})$ & $i{b}$  & Even \\
${A} \cos({\theta})$ & Odd & $({A}_{\text{odd}}, {\theta}_{\text{even}})$, $({A}_{\text{odd}}, {\theta}_{\text{odd}})$ & ${a}$ & Odd \\
$i {A} \sin({\theta})$ & Even & $({A}_{\text{even}}, {\theta}_{\text{even}})$,  $({A}_{\text{odd}}, {\theta}_{\text{odd}})$ & ${a}$ & Even \\
$i {A} \sin({\theta})$ & Odd & $({A}_{\text{even}}, {\theta}_{\text{odd}})$, $({A}_{\text{odd}}, {\theta}_{\text{even}})$ & $i{b}$ & Odd \\ 
\hline
\vspace{0.5mm}
\label{tab:FraunhoferSymmetries}
\end{tabular}
\end{table*}
\noindent Before we continue with an example of even phase aberrations, recall: 
\begin{itemize}
\item That a function $f(r)$ can be decomposed into even and odd functions:
\begin{align}
{f}({r})                       &= {f}_{\text{even}}({r}) + {f}_{\text{odd}}({r}), \\
{f}_{\text{even}}({r}) &= \frac{{f}({r})  + {f}(-{r})}{2}, \\
{f}_{\text{odd}}({r})   &= \frac{{f}({r})  - {f}(-{r})}{2}.
\end{align}
An example of such a symmetry decomposition of phase and amplitude in the context of the APP coronagraph is shown in \autoref{fig:APP_symmetries}. 
\item The multiplication and composition properties of even and odd functions: 
\begin{align}
f_{\text{even}}(r) \cdot g_{\text{odd}}(r) &= h_{\text{odd}}(r), \\
f_{\text{odd}}(r) \cdot g_{\text{odd}}(r) &= h_{\text{even}}(r), \\ 
f_{\text{even}}(r) \cdot g_{\text{even}}(r) &= h_{\text{even}}(r), \\
f_{\text{even}}[ g_{\text{odd}}(r)] &= h_{\text{even}}(r), \\
f_{\text{odd}}[  g_{\text{odd}}(r)] &= h_{\text{odd}}(r), \\ 
f_{\text{odd}}[  g_{\text{even}}(r)] &= h_{\text{even}}(r), \\
f_{\text{even}}[  g_{\text{even}}(r)] &= h_{\text{even}}(r).
\end{align}
\item The symmetry properties of Fraunhofer propagation, which are shown in \autoref{tab:FraunhoferSymmetries}.
\item The Hermitian properties of the Fraunhofer propagation. These say that a conjugated pupil-plane electric field ${E}_{\text{pup}}' = {E}_{\text{pup}}^*$ (i.e. a phase sign flip; $^*$ denotes the conjugation) will result in a flipped and conjugated focal-plane electric field ${E}_{\text{foc}}'=\mathcal{C}\{ {E}_{\text{pup}}' \}$:
\begin{equation}\label{eq:hermitianprop}
{E}_{\text{foc}}'({r}) = -{E}_{\text{foc}}(-{r})^*.     
\end{equation} 
\end{itemize}
The reason that the symmetry decomposition, combined with the decomposition of ${E}_{\text{pup}}$ in its real and imaginary components (\autoref{eq:pup_real_imag}), is important, is that the Fraunhofer propagation maps combinations of these decompositions into either real or imaginary components of ${E}_{\text{foc}}$, as shown in \autoref{tab:FraunhoferSymmetries}. It will be shown below that these properties of the Fraunhofer propagation determine what kind of symmetry has to be introduced in the pupil-plane to determine the sign of even phase aberrations. \\
\begin{figure*}
\centering
   \includegraphics[width=17cm]{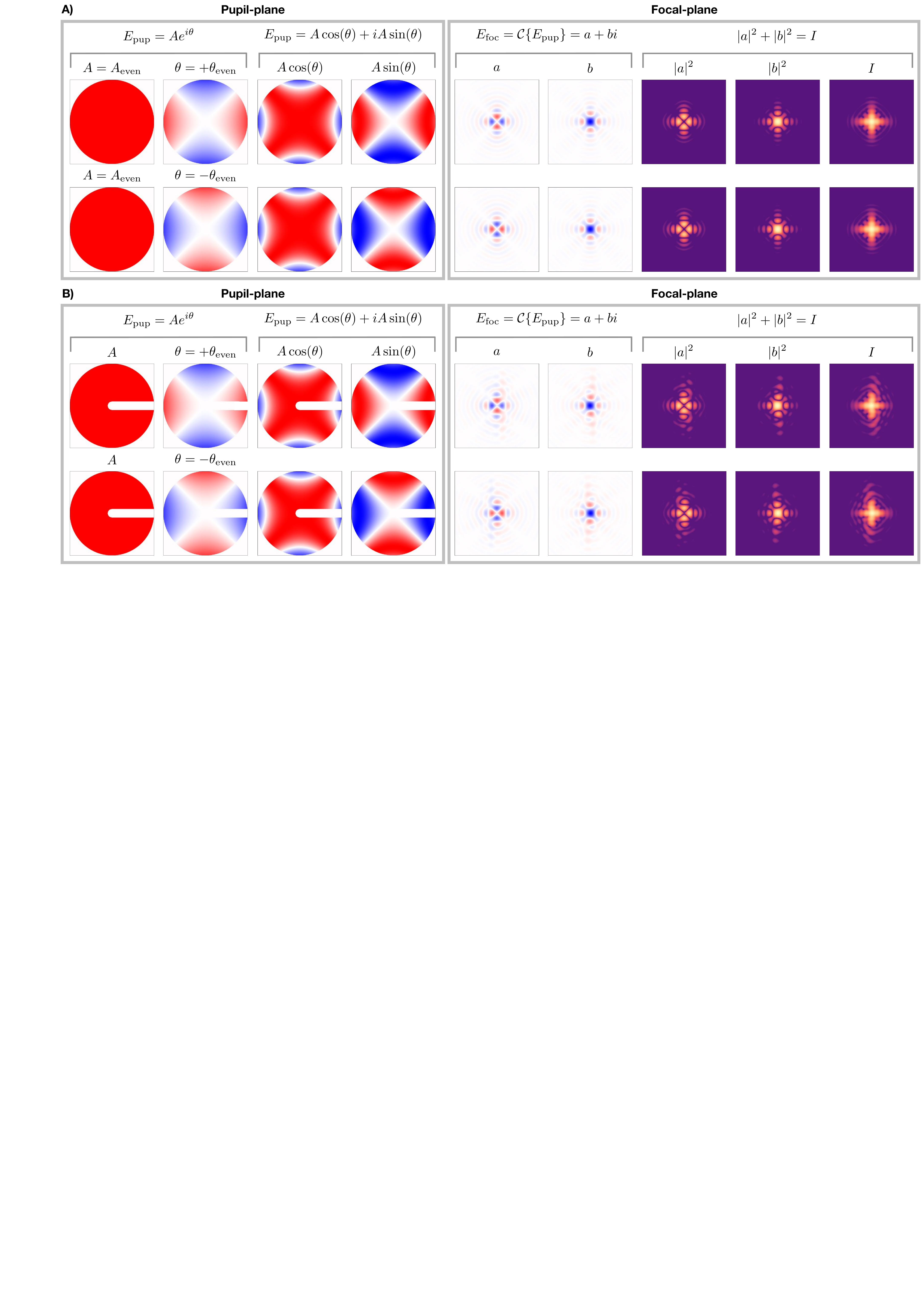}
     \caption{Focal- and pupil-plane quantities of an even phase aberration (astigmatism) through a (a) symmetric and (b) asymmetric pupil. The two rows show opposite signs for the phase aberration. The columns in the pupil-plane box show (from left to right) the amplitude, phase, real and imaginary electric field. In the focal-plane box, the columns show the real and imaginary electric fields, the power in the real and imaginary electric fields, and the total power.}
     \label{fig:theory_fig_A}
\end{figure*}
\indent Let's assume that $A$ is even, which is true for most instrument pupils. If an even phase aberration is added, e.g. astigmatism, the terms $A\cos(\theta)$ and $iA\sin(\theta)$ will both be even. Note that only the imaginary term contains sign information on the aberration. In this example, when they are propagated to the focal-plane, $A\cos(\theta)$ will go to $ib$ and $iA\sin(\theta)$ to $a$ (shown, respectively, in the first and third row of \autoref{tab:FraunhoferSymmetries}). These terms are still even and the sign information on the aberration is encoded in the real part of the focal-plane electric field. The resulting PSF recorded by the detector is even. These steps are shown in the top row of \autoref{fig:theory_fig_A} a. If the sign of the aberration flips (i.e a conjugation of the pupil-plane electric field), the PSF flips. But as the PSF is even, there is no morphology change recorded and thus the sign information cannot be retrieved. See the bottom row of \autoref{fig:theory_fig_A} a. One can only hope to determine the sign by measuring the real electric field, as that is where the sign information is encoded. For odd phase aberrations the sign flip will result in a morphology change and can therefore be measured. \\ 
\indent If the same exercise is performed with a pupil amplitude asymmetry, there will be a morphology change. This is shown in the top and bottom rows of \autoref{fig:theory_fig_A} b. The reason is that an even pupil amplitude, as shown in the top row of \autoref{fig:theory_fig_B} a, will only generate an imaginary electric field and therefore will not interfere with sign information containing real electric field caused by even aberrations. This is not the case for an asymmetric pupil amplitude, as it also generates a real electric field due to odd pupil amplitude (second row of \autoref{tab:FraunhoferSymmetries}). This real electric field will interfere with the aberration's real electric field and thus enable the sign determination. The electric field of an asymmetric pupil is shown in the bottom row of \autoref{fig:theory_fig_B} a. More examples on pupil-plane phase retrieval with focal-plane images can be found in \autoref{sec:phaseretrieval}. \\ 
  \indent This is the working principle of the Asymmetric Pupil Fourier Wavefront Sensor \citep{martinache2013asymmetric}. Interestingly, other focal-plane wavefront sensing methods such as the differential Optical Transfer Function \citep{codona2012experimental} and Self-Coherent Camera (\citealt{baudoz2005self}) also rely on pupil asymmetries, albeit use other reconstruction algorithms. Classical phase diversity techniques (\citealt{gonsalves1982phase};  \citealt{paxman1992joint}) come to a similar result by introducing a known, even phase aberration (e.g. defocus; odd phase modes can never be used, see the top row of \autoref{fig:theory_fig_B} b). This is the only other way of probing the real part of the focal-plane electric field (\autoref{tab:FraunhoferSymmetries}). Therefore, all these methods can now be understood as one family that probes the real focal-plane electric field by manipulating either pupil-plane phase or the pupil-plane amplitude. In the context of FPWFS with the vAPP, it is undesirable to use an even pupil-plane phase to break the sign ambiguity. This is because it fills up the dark-hole of the coronagraph and therefore prevents simultaneous science observations and wavefront measurements. On the other hand, vAPPs can be designed for pupils with an amplitude asymmetry, this will be shown in the next subsection, and therefore can combine science observations and wavefront sensing. This result also applies to other FPWFS techniques, for example see \autoref{sec:implications_LDFC} how it affects spatial LDFC. 

\subsection{vAPP design for phase retrieval}
Our framework of describing how pupil-plane phase and amplitude symmetries map to the focal-plane electric field also helps to understand some of the aspects of APP design and the requirements to turn an APP into a FPWFS. The optimization method that guarantees optimal APP designs, which turn out to be phase-only solutions, is detailed in \cite{por2017optimal} and will not be discussed here. The APPs considered in this section are all designed to have a one-sided D-shaped dark hole from 1.8 $\lambda/D$ to 10 $\lambda/D$ with a raw contrast of $<10^{-5}$. The dark hole is defined as the area in the focal-plane where the raw contrast meets the requirement. \\
\indent Suppose that we want to design an APP with a one-sided dark hole (= PSF with odd symmetry) for a symmetric aperture ($A_{\text{even}}$). Such an aperture will yield a completely imaginary focal-plane electric field with even symmetry as shown in the top row of \autoref{fig:theory_fig_B}. As the APP manipulates phase, this imaginary electric field needs to be cancelled on one side using pupil-plane phase. The last row of \autoref{tab:FraunhoferSymmetries} indicates that the only way of removing an odd part of the imaginary focal-plane electric field ($ib$) for an even aperture ($A_{\text{even}}$) is by introducing a purely odd pupil-plane phase. This indeed results from the optimization method, see the top row of  \autoref{fig:theory_fig_B} b, and \autoref{fig:APP_symmetries} a.\\
\indent An asymmetric aperture, e.g. as shown in the bottom row of \autoref{fig:theory_fig_B} a, yields a focal-plane electric field with real and imaginary components. The asymmetric aperture consists of a central obscuration that is $10\%$ of the aperture diameter and a bar with the same width that connects the central obscuration with the edge of the aperture. Cancelling such a focal-plane electric field on one side to create the dark hole requires an odd pupil phase to cancel the imaginary focal-plane electric field (last row of \autoref{tab:FraunhoferSymmetries}), and an even pupil-plane phase for the real part (third row of \autoref{tab:FraunhoferSymmetries}). Again, this is indeed the solution the optimization method comes up with, as shown in the bottom rows of \autoref{fig:theory_fig_B} b, and \autoref{fig:APP_symmetries} b.  Crucially for FPWFS with the APP, the real focal-plane electric field originating from the \textit{odd} pupil amplitude component cannot be completely removed by the \textit{even} pupil-plane phase, but is enhanced on the bright side of the APP coronagraphic PSF, as shown in the bottom row of \autoref{fig:theory_fig_B} b. \\ 
\indent To summarize, FPWFS capabilities of the vAPP are fundamentally enabled by the pupil amplitude asymmetry. This asymmetry introduces a real focal-plane electric field that interferes with the even pupil-plane phase aberrations and enables their sign determination. The vAPP allows for simultaneous science observations by removing the the real electric field from the dark hole and moving it to the bright field. See \autoref{sec:phaseretrieval} for figures that demonstrate this.
\section{Aberration estimation algorithm}\label{sec:model} 

\subsection{Maximum a posteriori estimation}
We have developed an algorithm that gives a maximum a posteriori (MAP) estimation of the phase aberrations by maximizing the posterior likelihood $p(\alpha, N_p, N_b, v, L |{D})$. It takes into account a physical, non-linear model of the vAPP (shown in the next subsection), noise statistics and prior knowledge of the estimated parameters. Given an image ${D}$ the algorithm estimates $\alpha$, a vector containing the amplitudes of the phase aberration modes of interest. It has the option to additionally estimate the following parameters: the number of source photons in the image $N_p$,  the background level $N_b$, the fractional degree of circular polarization $v$, and the fractional strength of the leakage PSF integrated over the spectral band $L$. These parameters are summarized in \autoref{tab:algorithm_variables}. The estimation of the phase aberrations is more conveniently performed by minimizing $\mathcal{L}(\alpha, N_p, N_b, v, L| {D})$; the negative logarithm of the likelihood function ($\mathcal{L}(\alpha, N_p, N_b, v, L| {D}) = - \ln[p(\alpha, N_p, N_b, v, L |{D})]$: 
\begin{equation}
\hat{\alpha}, \hat{N_p}, \hat{N_b}, \hat{v}, \hat{L} = \arg \min_{\alpha, N_p, N_b, v, L} \mathcal{L} (\alpha, N_p, N_b, v, L| {D}). 
\end{equation}
Parameters of $\mathcal{L}$ that have been estimated are denoted with a hat. The objective function $\mathcal{L}$ is given by: 
\begin{equation}\label{eq:objective_function}
\mathcal{L} (\alpha, N_p, N_b, v, L| {D}) = \sum_{{x}} \frac{1}{2 {\sigma}_n^2} ({D} - {M}(\alpha, N_p, N_b, v, L))^2 + \mathcal{R}(\alpha)
\end{equation}
with ${D}(x)$ the 2D image containing the data and ${M}(\alpha, N_p, N_b, V, L)$ the 2D model of the vAPP PSF which will be detailed in the next subsection. The algorithm fits the model to the entire image and thus does not exclude any regions in $D(x)$. The sum over ${x}$ is over all spatial pixels. The algorithm will be applied to long-exposure images that have high photon numbers, which can be approximated to contribute a spatially changing Gaussian noise ($\sigma_p^2$), and detector noise, assumed to be white Gaussian noise  ($\sigma_d^2$). Therefore, the total noise is Gaussian and the variance is the sum of the two independent processes variances ($\sigma_n^2 = \sigma_p^2 + \sigma_d^2$). Prior information on the phase aberrations is taken into account explicitly by adding a term $ \mathcal{R}(\alpha)$ to the objective function, which is given by:
\begin{equation}\label{eq:regularization}
\mathcal{R} ({\alpha}) = \frac{1}{2} \sum_{k=1}^N \frac{ \alpha_k^2 } {k^{\gamma}}, 
\end{equation}
with $\alpha_k$ the modal coefficients and $\gamma$ the assumed power spectrum. This term penalizes high spatial frequency modes according to assumptions on or measurements of the spatial power spectrum. Implicitly, we also regularize by using a truncated mode basis. \\
\indent We minimize $\mathcal{L}$ using the bounded limited-memory Broyden-Fletcher-Goldfarb-Shanno algorithm (L-BFGS-B; \citealt{byrd1995limited}), which is a quasi-Newton optimization algorithm that assumes a differentiable scalar objective function. The algorithm accepts analytically calculated gradients to increase the speed and accuracy. Such analytical expressions of \autoref{eq:objective_function}'s gradients 
 are given in \autoref{sec:derivatives}. The algorithm also requires a start position for the estimated parameters. Generally, $N_p, N_b, v$ and $L$ can be easily estimated from the data, as will be shown in \autoref{sec:demonstration}. From experience, setting $\alpha = 0$ (i.e. no aberration) works best when there is no initial guess available for the aberrated wavefront. 
\begin{table}
\caption{Parameters presented in \autoref{sec:model}.}
\label{tab:algorithm_variables}
\vspace{2.5mm}
\centering
\begin{tabular}{l|l}
\hline
\hline
Variable & Description \\ \hline
$\alpha$ & Vector containing the modal coefficients $\alpha_i$.\\ 
$\hat{\alpha}$ & Estimation of $\alpha$ by algorithm. \\
$\gamma$ & Assumed power spectrum in the regularization term.\\
$\theta_j$ & Pupil-plane phase of PSF $j$ (aberration + vAPP). \\
$\theta_{\text{APP}}$ & Pupil-plane phase of the vAPP. \\
$\sigma_d^2$ & Variance of detector noise.\\
$\sigma_n^2$ & Total noise variance ($\sigma_n^2$ = $\sigma_p^2$ + $\sigma_d^2$). \\
$\sigma_p^2$ & Variance of photon noise. \\
$\phi_i$ & Pupil-plane phase of aberration mode $i$. \\
$a_j$ & PSF $j$'s relative intensity, see \Crefrange{eq:a_1}{eq:a_3}.\\
$A$ & Pupil-plane amplitude. \\
$\mathcal{C}\{\ \cdot \}$ & Fraunhofer propagator \citep{goodman2005introduction}. \\
$E_{\text{pup}}$ & Pupil-plane electric field. \\
$D$ & Image used for wavefront sensing. \\
$I_{\text{foc},j}$ & PSF $j$, see \autoref{eq:focal_intensity}.\\
$L$ & Fractional strength of leakage PSF.\\
$\hat{L}$ & Estimation of $L$ by algorithm. \\
$\mathcal{L}$ & Objective function, see \autoref{eq:objective_function}. \\
$M$ & PSF model of vAPP, see \autoref{eq:PSF_model}.\\
$N_p$ & Total number of photons in image.\\
$\hat{N_p}$ & Estimation of $N_p$ by algorithm. \\
$N_b$ & Background level in image. \\
$\hat{N_b}$ & Estimation of $N_b$ by algorithm. \\
$\mathcal{R}$ & Regularization term, see \autoref{eq:regularization}.\\
$v$ & The fractional degree of circular polarization. \\
$\hat{v}$ & Estimation of $v$ by algorithm. \\
\hline
\end{tabular}
\end{table}
\subsection{Coronagraph model}
The vAPP is a half-wave retarder with a spatially varying fast-axis; these spatial variations induce geometric phase on the circular polarization states. Opposite circular polarizations receive the opposite phase, creating two similar, but mirror-imaged coronagraphic PSFs, see \autoref{fig:gvAPP_example}. Additionally, as the optic will not be perfectly half-wave, part of the light will not receive the desired phase and will create an extra non-coronagraphic PSF; we will refer to this PSF as the leakage. The relative intensity of the coronagraphic PSFs depends on the fractional degree of circular polarization $v$ (normalized to the total intensity) and the amount of leakage. In the model we assume a single point source and therefore the image plane is given by:
\begin{equation}\label{eq:PSF_model}
{M}({\alpha}, N_p, N_b, v, L) = N_p \left(\sum_{j=1}^3 a_j (v, L) {I}_{\text{foc}, j} ({\alpha}) \right) + N_b, 
\end{equation} 
with $N_p$ the total number of photons in the image, $N_b$ the background level (e.g. sky background, dark current or bias), ${I}_{\text{foc}, j} $ the image from by either left- ($j$=1) or right-handed ($j=$2) circular polarization or leakage ($j$=3). The $a_j(v, L)$ terms describe the relative intensities of these PSFs and are given by:
\begin{align}
\label{eq:a_1} a_1(v, L) &= \frac{1+v}{2} (1-L), \\ 
\label{eq:a_2} a_2(v, L) &= \frac{1-v}{2} (1-L), \\ 
\label{eq:a_3} a_3(v, L) &= L. 
\end{align}
PSFs are normalized such that the sum over all pixels in the image is one, i.e. $\sum_{{r}} {I}_{\text{foc}, j} = 1$. Note that we can incoherently add the leakage term and the coronagraphic PSFs, because orthogonal polarization states emitted by blackbody sources are incoherent, even when they are transformed to the same polarization state (Fresnel-Arago laws; \citealt{kanseri2008observation}; \citealt{mujat2004law}). The ${I}_{\text{foc}, j} $ is given by:
\begin{align}\label{eq:focal_intensity}
{I}_{\text{foc}, j} ({\alpha})&= | \mathcal{C} \{ {E}_{\text{pup}} (\alpha)\}|^2 \\
				       &= | \mathcal{C} \{ {A} e^{i ({\theta}_j + \sum_i \alpha_i {\phi}_i)} \}|^2 
\end{align}
with ${A}$ the amplitude of the aperture, ${\theta}_j = \pm {\theta}_{app}$ for $j=1,2$ and ${\theta}_j = 0$ for $j=3$. The phase aberration is expanded in the mode basis $\{{\phi}_i \}$, with $\alpha_i$ being the estimated coefficients. This last equation shows the non-linearity of the model: 1) we estimate phase, which is contained in the complex exponent, and 2) the estimation is performed on intensity images, which are the square of the electric field. 
\begin{figure*}
\centering
   \includegraphics[width=17cm]{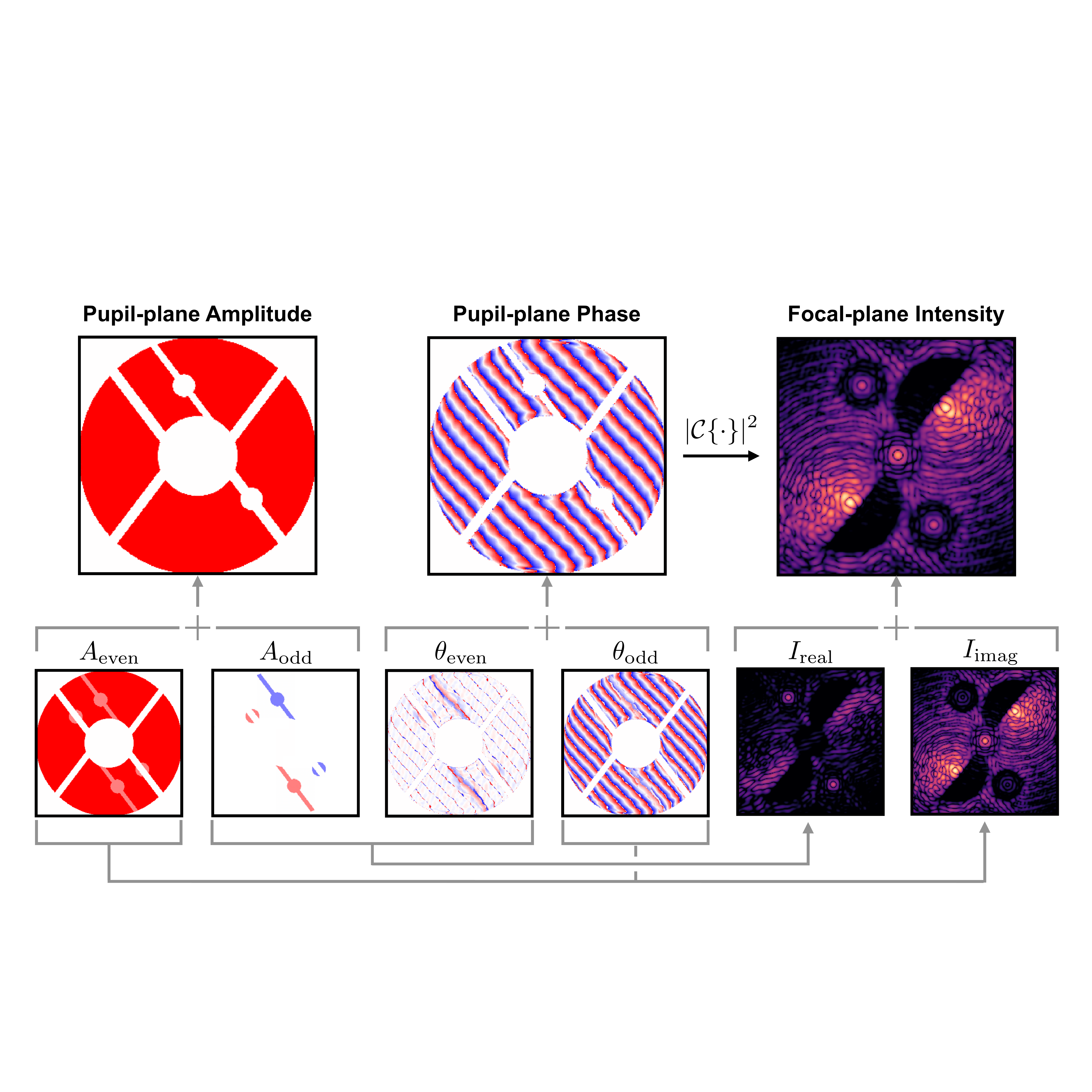}
     \caption{Amplitude and phase design of the SCExAO vAPP \citep{doelman2017patterned} and its resulting PSF. The pupil-plane quantities are decomposed into their even and odd constituents, while the focal-plane PSF is decomposed into the power in the real and imaginary part of the electric field. }
     \label{fig:power_scexao_vapp}
\end{figure*}
\section{Simulations} \label{sec:simulations}
In this section we explore the performance of the algorithm in idealised circumstances by numerical simulations. Performance will be reported in absolute units (nm) in the context of the SCExAO vAPP and in relative units (fractional $\lambda$) to allow for comparison with other methods. \\
\indent The design of the SCExAO vAPP is detailed in \cite{doelman2017patterned}; here we only discuss the most relevant details. The D-shaped dark holes have a raw contrast of $10^{-5}$ w.r.t. the peak flux of the coronagraphic PSF and extend from 2 to 11 $\lambda/D$. Besides the two coronagraphic PSFs, and the central leakage PSF, two additional phase diversity PSFs were added. These PSFs have opposite defocus aberration and can be used for classical phase diversity, but are not considered in this work. The top row of \autoref{fig:power_scexao_vapp} shows the pupil-plane amplitude for which the vAPP was designed, the subsequent phase design and the focal-plane intensity. As detailed in the bottom row of the figure, there is a clear amplitude asymmetry due to the blocking of dead DM actuators. This odd amplitude component ${A}_{\text{odd}}$, combined with the even phase $\theta_{\text{even}}$ that cancels the effect of the asymmetry in the dark holes, results in the real focal-plane electric field that will be used to probe even pupil-plane phase aberrations. Approximately $3.5$ \% of the focal plane power is in the real electric field and therefore available to sense even phase aberrations. \\
\indent Simulations are performed with the HCIPy package \citep{por2018hcipy}, an open-source software for high-contrast imaging simulations that is available on GitHub\footnote{\url{https://github.com/ehpor/hcipy}}. The simulations are monochromatic ($\lambda$=1550 nm). They sample the pupil-plane with 256$\times$256 pixels and use the Fraunhofer approximation to propagate the electric field to the focal-plane, which is sampled with 150$\times$150 pixels for 3 pixels per $\lambda/D$. The aberration estimation algorithm is Python-based, uses the HCIPy package for the coronagraph model and the Scipy library \citep{jones2014scipy} to minimize the objective function.
\begin{figure}
\centering
\includegraphics[width=\hsize]{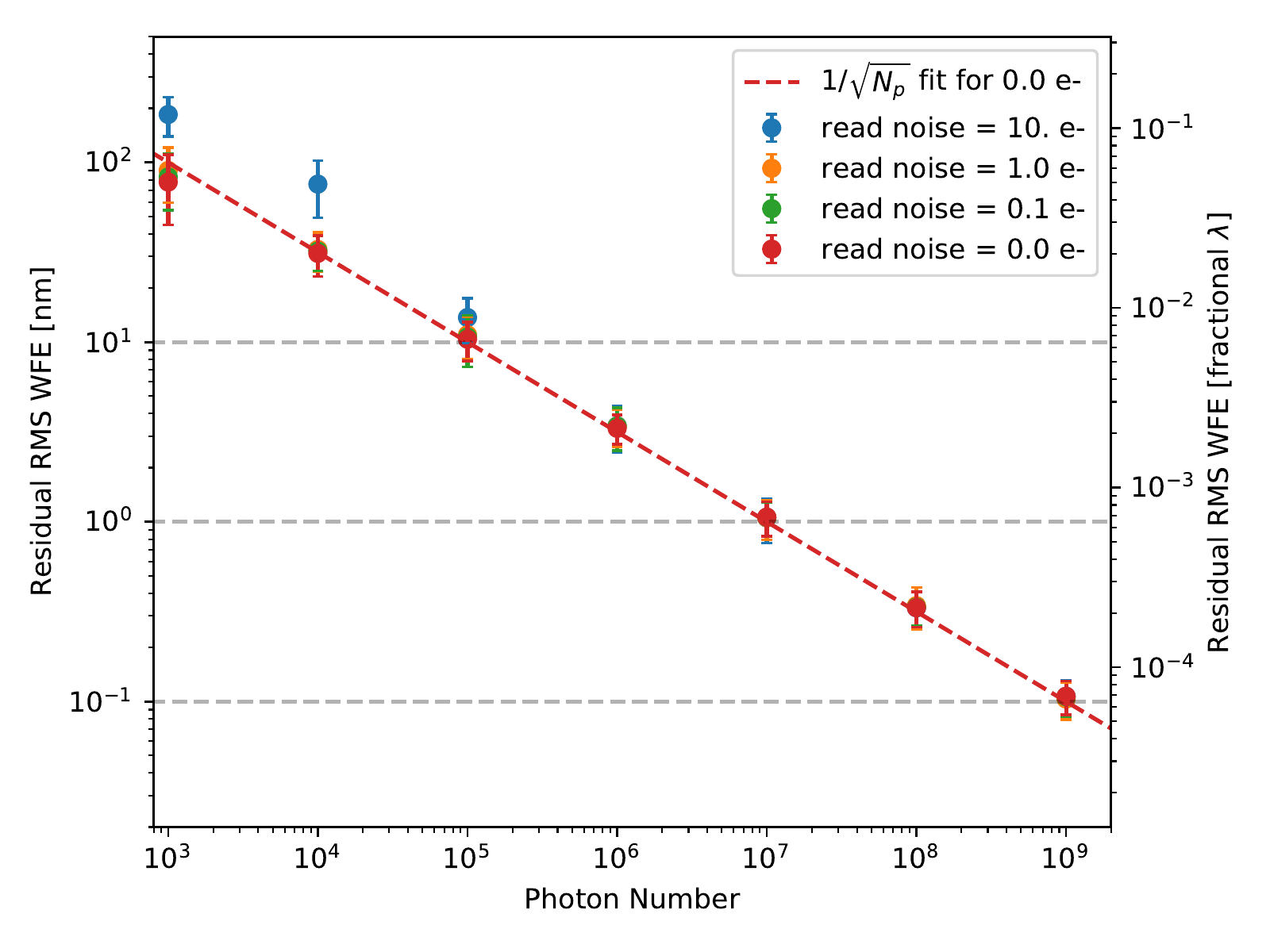}
\caption{Residual RMS WFE as a function of photon number $N_p$ for the SCExAO vAPP. Different colours denote various read noise levels. The dotted line shows a $1/\sqrt{N_p}$ fit to the pure photon noise case. The RMS WFE in nm assumes $\lambda = 1550$ nm.}
\label{fig:photon_noise_sens}
\end{figure}
\subsection{Photon and read noise sensitivity}
\autoref{fig:photon_noise_sens} presents the sensitivity of the algorithm and the SCExAO vAPP when considering varying levels of photon and read noise. The modal basis set used for these tests consisted of the first 30 Zernike modes, starting at defocus. All induced aberrations were a linear sum of these modes. The aberrated phase screen is created by first calculating a phase screen following a spatial power spectrum with a slope of -2.5. This slope is steeper than a slope of -2, usually considered for the power spectrum of NCPA, but was chosen to put more power in the low-order modes as we focus on measuring those. Subsequently, this phase screen is projected onto the 30 Zernike modes, which were added together and scaled to a $\sim\lambda/16$ (100 nm) RMS WFE. The algorithm estimated the same Zernike modes as were used to generate the phase screen. For every photon number $N_p$ ($10^3 - 10^9$ photons) we generated 10 different phase screens and, for each phase screen, 10 photon noise realizations. Furthermore, four levels of read noise were tested: $0$ e$^-$, $0.1$ e$^-$, $1$ e$^-$ and $10$ e$^-$. The results of the simulations are plotted in \autoref{fig:photon_noise_sens}, where the dots denote the mean RMS WFE per photon number and read noise level, and the error bars denote the $1 \sigma$ deviation. The dotted line is a fit of a $\propto 1 / \sqrt{N_p}$ function to the pure photon noise simulations, confirming that the performance scales with the photon noise. The algorithm therefore is photon noise limited, i.e. no algorithmic effects limit the performance. For the photon-limited case, a $<\lambda/1000$ residual RMS WFE is reached with $10^7$ photons. In the context of the SCExAO system (\autoref{sec:demonstration}) this would be a 1-nm residual WFE for a $\sim$5 second integration on a $m_H=8$ star. The simulations also show that increasing the read noise to 0.1 e$^-$ or 1 e$^-$ does not significantly impact the performance of the algorithm, i.e. the error is still dominated by photon noise. For a read noise level of 10 e$^-$ and photon numbers $< 10^6$, the read noise starts to dominate the error and decreases the performance by a factor $\sim 2 - 3$.
\begin{figure}
\centering
\includegraphics[width=\hsize]{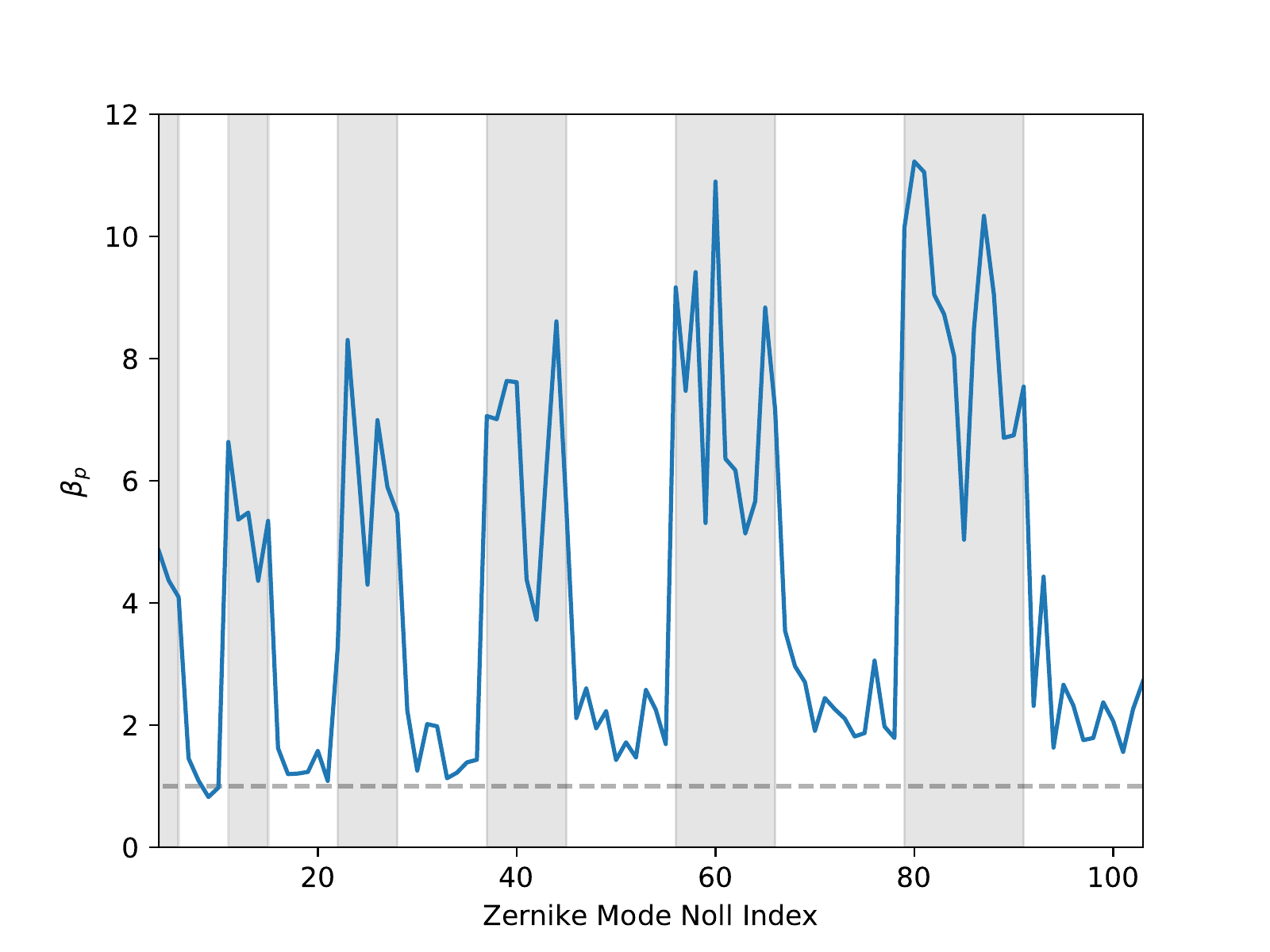}
\caption{Photon noise sensitivity of the SCExAO vAPP to the first 100 Zernike modes, starting at defocus. The grey boxes denote the even Zernike modes, and the white boxes the odd Zernike modes.}
\label{fig:mode_sens}
\end{figure}
\subsection{Mode photon noise sensitivity}
In \cite{guyon2005limits} the photon noise sensitivity $\beta_p$ of a WFS to a single mode $m$ is defined as: 
\begin{equation}
\beta_p = \sigma_{m} \sqrt{N_p},
\end{equation}
with $\sigma_m$ the reconstruction error for mode $m$ and $N_p$ the number of photons in the image. According to \cite{guyon2005limits}, the ideal WFS will have $\beta_p = 1$ for all reconstructed modes. Here we explore the sensitivity of the SCExAO vAPP to the first 100 Zernike modes (starting with defocus as the vAPP is tip/tilt insensitive), expecting, as discussed in \autoref{sec:theory} that the vAPP has a lower sensitivity (resulting in a higher $\beta_p$) for the even modes than for the odd modes. The reconstruction error $\sigma_m$ was determined by calculating the standard deviation of the $\hat{\alpha}_m$ distribution, obtained after simulating 100 different photon noise realisations per photon number ($10^7$, $10^8$ and $10^9$) and estimating only $\hat{\alpha}_m$ (no other modes were estimated simultaneously). This resulted in three $\beta_p$'s (for every tested photon number), which were subsequently averaged to obtain the final result. The results are shown in \autoref{fig:mode_sens}, with the ideal WFS ($\beta_p = 1$) shown as the horizontal grey dotted line. The grey boxes denote the even modes, and the white boxes the odd modes. The first thing to note is that the even modes perform a factor $\sim 4 - 10$ worse than the odd modes. This is because, as discussed in \autoref{sec:theory}, the real part of the focal-plane electric field probes the even modes, and is in the case of the SCExAO vAPP is relatively weak: $\sim$3.5\% of the total power is in the real electric field. Therefore, it is expected that $\beta_p$ is $\sqrt{1/0.035} = 5.3$ times higher for even modes compared to the odd modes, which is approximately observed. Furthermore, it should be noted that the spatial frequency content of the Zernike modes increases with Noll index. Therefore, due to lower SNR at higher frequencies, sensitivity to both even and odd modes decreased with increasing mode number. The sensitivity to even modes decreases faster than the odd mode sensitivity, this is because the real part of the focal-plane electric field is limited in extent as compared to the imaginary part (shown in \autoref{fig:power_scexao_vapp}). Note that this analysis does not include any cross-talk effects, which will be taken into account in the next section. 
\begin{figure}
\centering
\includegraphics[width=\hsize]{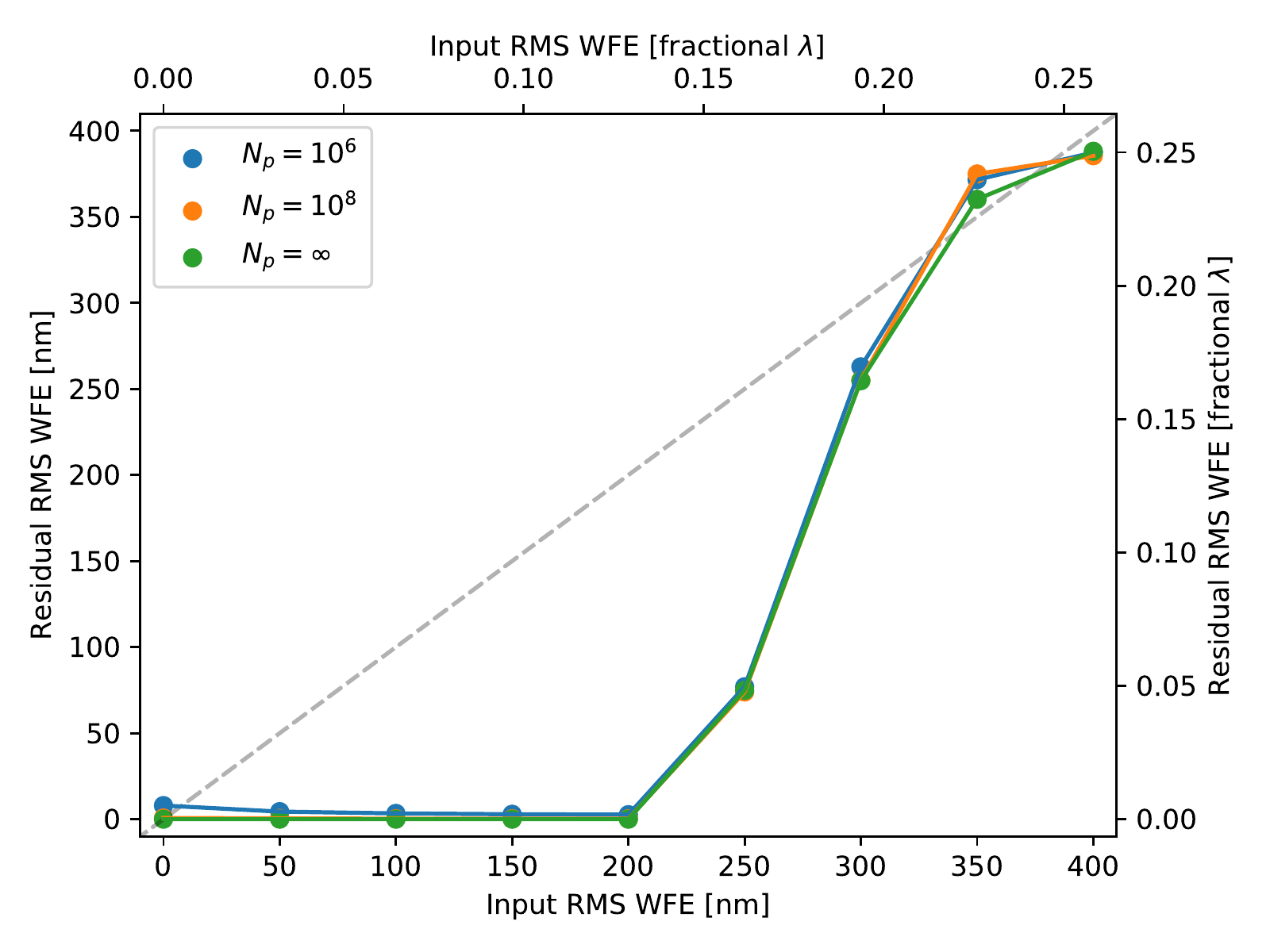}
\caption{Residual RMS WFE as function of introduced RMS WFE for the SCExAO vAPP. The colours denote various photon numbers. The dashed, grey line separates the regions where the wavefront improves or deteriorates. The wavelength is 1550 nm.}
\label{fig:wfe_sens}
\end{figure}
\subsection{Dynamic range algorithm}
Here we explore the maximum RMS WFE for which the algorithm can accurately recover estimates for each modal coefficient in a single step. This limitation is mainly driven by modal crosstalk and will determine the WFE for which the algorithm can converge to a low ($\sim\lambda/1000 - \lambda/100$ or 1-10 nm) residual RMS WFE in one iteration, assuming that the estimation is perfectly applied by the DM. We simulated phase screens consisting of 30 Zernike modes (starting with defocus), following a spatial power spectrum with a $-2.5$ slope, with increasing WFE. For RMS WFE values between 0 and $\sim \lambda/4$ (400 nm) we simulated 10 wavefronts, with either no photon noise ($N_p = \infty$) or high photon numbers ($N_p = 10^6, 10^8$). For the simulations with photon noise, we simulated 10 photon-noise realisations per phase screen. The algorithm estimated the same Zernike modes. The residual RMS WFE was determined by subtracting the estimation from the initial, aberrated wavefront. The mean residual RMS WFEs are shown in \autoref{fig:wfe_sens}. The performance of the algorithm does not significantly vary between the different photon number realisations. The algorithm converges in one iteration with an input aberration RMS of up to $\sim \lambda/8$ or 200 nm. For WFE between $\lambda/8$-$\lambda/5$ (200-300 nm), the algorithm still improves the wavefront, but there will be significant residual WFE after one iteration. In this regime, the algorithm will converge to $< \lambda/1000$ or 1-nm RMS WFE within two or three iterations in closed-loop operation. A WFE larger than $\lambda/5$ (300 nm) RMS does not significantly improve after one iteration. Note that this performance is very similar to the non-coronagraphic results reported by \cite{martinache2016closed} for the Asymmetric Pupil Fourier Wavefront Sensor.
\section{Demonstration at SCExAO}\label{sec:demonstration}
\subsection{SCExAO}
The Subaru Coronagraphic Extreme Adaptive Optics (SCExAO) instrument \citep{jovanovic2015subaru} is a high-contrast imaging instrument at the Subaru Telescope. It operates downstream of the AO188 system \citep{minowa2010performance}, which gives an initial, low-order correction. SCExAO drives a 2000-actuator deformable mirror (DM) based on wavefront measurements from a pyramid wavefront sensor (PYWFS) in the optical - NIR (600-900 nm) that can run at 3.5 kHz, but most often runs at 2 kHz. There are 45 actuators across the pupil, giving the system the ability to correct out to $22.5 \lambda / D$. The real-time control of SCExAO is handled by the compute and control for adaptive optics (cacao) package \citep{guyon2018compute}. Cacao allows for NCPA corrections to be handled by a separate DM channel, where the software automatically sends offsets to the PYWFS. These offsets ensure that the PYWFS does not sense and attempt to control the NCPA corrections. The instrument routinely achieves Strehl ratios $> 80 \%$ in H-band on-sky. SCExAO feeds the post-AO JHK-bands to the integral field spectrograph CHARIS (\citealt{peters2013optical}; \citealt{groff2014construction}). The ultimate goal is to operate the vAPP FPWFS with the CHARIS data; however, the tests were performed with the internal NIR camera for ease of operation. The NIR camera has recently \citep{lozi2018scexao} been upgraded to a C-RED 2 detector \citep{feautrier2017c}, which has a 640$\times$512 pixel InGaAs sensor that is cooled to $-40$ $^{\circ}$C. 
\subsection{Algorithm implementation in SCExAO}
The algorithm has to be adapted to the SCExAO parameters and the vAPP design. To calibrate the model of the vAPP we had to take into account the following:
\begin{itemize}
\item \textbf{Pupil undersizing}: to account for pupil misalignments the design of the vAPP was undersized by $1.5\%$ \citep{doelman2017patterned} w.r.t. the nominal SCExAO pupil.
\item \textbf{Pupil rotation}: to optimally block dead actuators in the DM, the SCExAO pupil was rotated by $-6.25^{\circ}$, affecting the orientation of the PSF on the detector. 
\item \textbf{Detector pixel scale}: for accurate propagation to the focal-plane of the NIR camera the pixel scale was determined manually (which was deemed sufficient for low-order Zernike mode measurement and correction) to be $15.6$ mas per pixel at 1550 nm.
\end{itemize}
The coronagraph model described in \autoref{sec:model} assumes a monochromatic light source, and the closest available option, in terms of bandwidth, for the NIR camera in SCExAO is the $\Delta \lambda = 25$ nm filter centered around $\lambda = 1550$ nm. For this filter, a $128\times128$ pixel subwindow in the NIR internal camera is sufficient to contain the vAPP PSFs. The image fed to the algorithm consists of the sum of a 1000 unsaturated images (either of the internal source or the star; individual exposure times are on the order of 1-10 milliseconds) that are dark-subtracted, aligned with a reference PSF generated with the coronagraph model and subsequently stacked and calibrated for the camera system gain (2.33 e$^-$/ADU; \citealt{feautrier2017c}). The large number of stacked images should generally bring the algorithm into the 1-nm residual RMS WFE regime (\autoref{fig:photon_noise_sens}) in ideal circumstances and is the best performance we can expect. After image acquisition and before running the algorithm, the parameters $N_p, N_b, v$ and $L$ have to be estimated:
\begin{itemize}
\item $N_b$ is estimated first by selecting the corners in the image where there is no light from the vAPP.
\item $N_p$ is estimated by subtracting $N_b$ from the image and summing the values in all pixels.
\item The grating in the vAPP acts as a circularly polarized beam-splitter, thus $v$ is determined by aperture photometry on the two coronagraphic PSFs. From these intensity measurements, $I_1$ and $I_2$, the fractional degree of circular polarization can be determined by:  
\begin{equation}\label{eq:v_determination}
v = \frac{I_1 - I_2}{I_1 + I_2}
\end{equation}  
\item Similarly, the amount of leakage $L$ is calculated by aperture photometry on the two coronagraphic ($I_1$ and $I_2$) and the leakage PSFs ($I_3$):
\begin{equation}\label{eq:L_determination}
L = \frac{I_{3}}{I_1 + I_2 + I_3}
\end{equation}
\end{itemize} 
After this initial estimation, the estimation of these parameters can also be improved by the algorithm. This is generally not done for $N_b$, $v$ and $L$ as it has not been found to improve the wavefront estimation. On the other hand, we have found that estimating $N_p$ using the algorithm does improve the estimation. Another concern is that estimating some of these parameters could induce cross-talk with the aberration coefficients. An example of such cross-talk would be between the wavefront aberration coma and $v$, as both parameters result in relative brightness changes between the two coronagraphic PSFs. Therefore, the algorithm could tune either parameter to fit a relative brightness change and arrive at the wrong result. We did not extensively study the effects of this cross-talk. \\
\indent The closed-loop correction is performed by phase conjugation, where the DM command $\theta_{\text{DM}, i}$ at iteration $i$ is calculated as:
\begin{equation}
\theta_{\text{DM}, i} = \theta_{\text{DM}, i-1} - \frac{g}{2} \sum_j \alpha_j \phi_j,
\end{equation}
with $g$ the closed-loop gain that can be freely chosen, the factor $\frac{1}{2}$ to account for the reflective nature of the DM, $\alpha_j$ the estimated modal coefficients and $\{ \phi_j\}$ the mode basis. \\
\indent When estimating 30 Zernike modes with the SCExAO implementation of the algorithm, the computation time per iteration is $\sim$$30-50$ seconds. Generally, the convergence time increased when sensing more modes and larger wavefront errors. The convergence time was mainly limited by the Python library versions installed at SCExAO, and more fundamentally by the implementation in Python. 
\begin{figure*}[ht]
\centering
   \includegraphics[width=17cm]{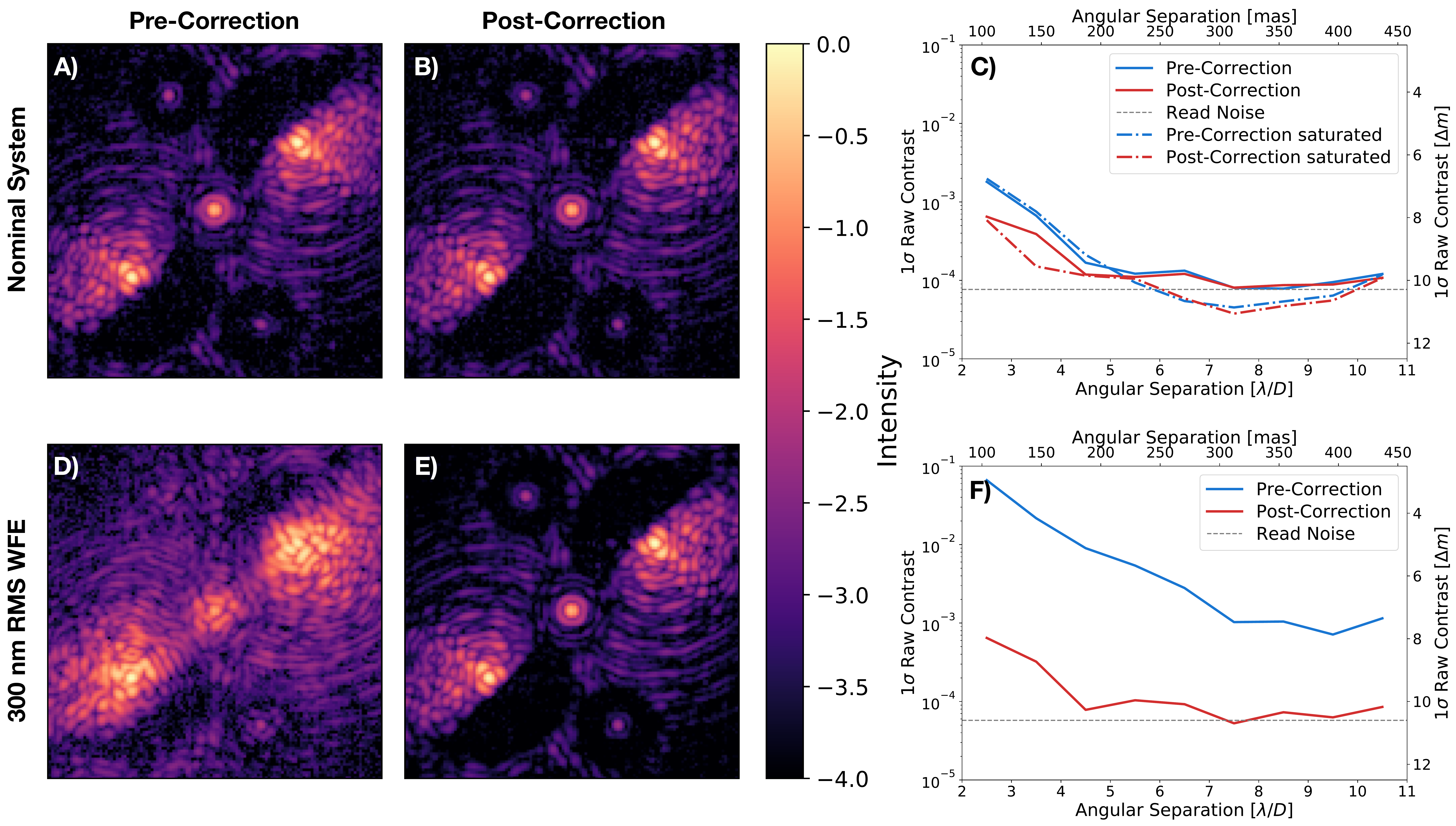}
     \caption{Focal-plane images and 1$\sigma$ raw contrast curves pre- and post-correction with the internal source within SCExAO. (a) Pre- and (b) post-correction images for the nominal SCExAO system. (c) The 1$\sigma$ raw contrast curves for the correction of the nominal system. The solid lines relate to subfigures (a) and (b) and are read noise limited beyond $7.5$ $\lambda/D$, which is shown by the horizontal dashed line. The dot-dashed lines show a similar correction performed at a different time. For these lines the images for raw contrast determination were saturated such that they were speckle noise limited (instead of read noise limited), reaching deeper contrasts in the dark holes. (d) Pre- and (e) post-correction images for an introduced wavefront of 300 nm RMS. (f) The 1$\sigma$ raw contrast curves for the correction of the system with the introduced 300 nm RMS wavefront, with again the horizontal dashed line denoting the read noise. }
     \label{fig:lab_results}
\end{figure*}
\begin{figure*}[ht]
\centering
   \includegraphics[width=17cm]{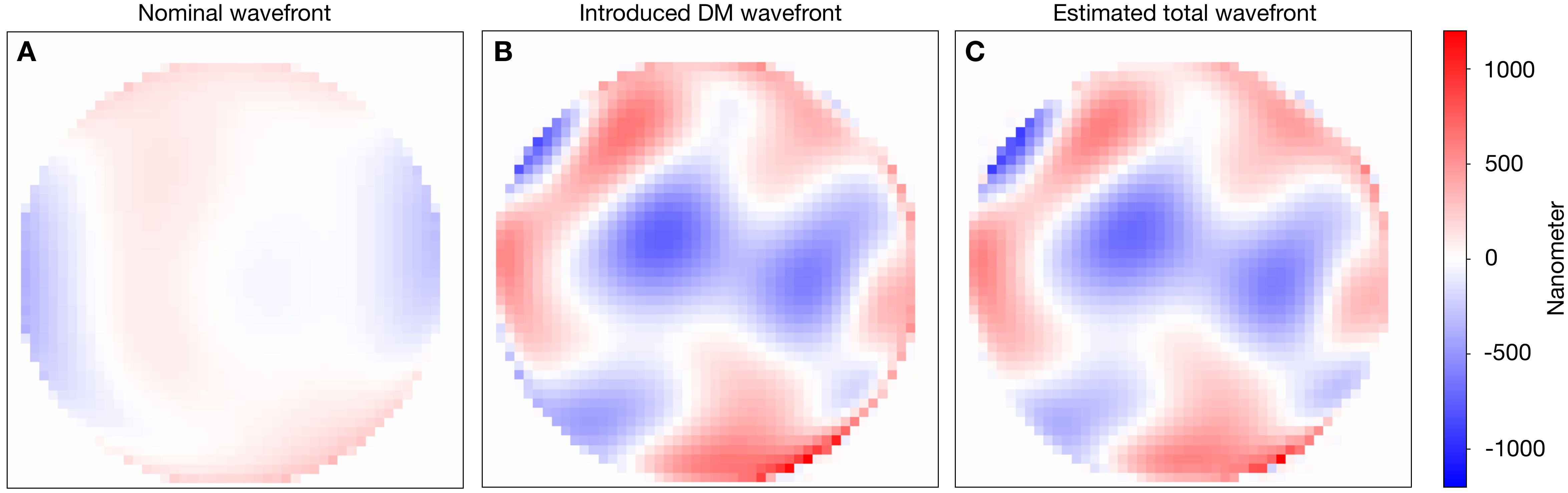}
     \caption{ (a) Estimated wavefront of the nominal SCExAO system using the internal source and reconstructed by combining the estimated modal coefficients of the 30 Zernike modes. The estimated WFE is $\sim$100 nm RMS. (b) The introduced and (c) estimated wavefront from closed-loop tests with the internal source. The introduced wavefront (b), consisting of 50 Zernike modes, has a RMS WFE of $\sim$293 nm. The estimated wavefront (c), using the same 50 Zernike modes, with an RMS WFE of $\sim$311 nm. Note that the estimated wavefront also includes other WFE already present in the system.}
     \label{fig:wavefront_aberrated_internal}
\end{figure*}
\subsection{Internal source demonstration}\label{sec:internal_source} 
We performed FPWFS tests with the SCExAO vAPP using the internal source in November and December 2018. The goal was to demonstrate the principle of FPWFS with the vAPP coronagraphic PSFs, and we therefore focussed on measuring and correcting low-order modes (30-50 Zernike modes, starting with defocus). \autoref{fig:lab_results} shows these results, where the 1$\sigma$ raw contrast was determined by calculating the standard deviation of 1 $\lambda / D$ wide annuli covering both PSFs that were normalized by the maximum number of counts in the image. The top row in the figure presents the pre- and post-correction PSFs and the radial averaged contrast improvements (the subfigures a, b and c, respectively) for the nominal SCExAO system using 30 Zernike modes, after one iteration with the algorithm ($g=1$). The estimated wavefront is shown in \autoref{fig:wavefront_aberrated_internal} a, and the measured WFE in the 30 Zernike modes is $\sim$100 nm RMS. The dominant mode is astigmatism at $\sim$86 nm RMS. The next three largest aberrations were coma, quadrafoil and secondary astigmatism at respectively $\sim$28 nm, $\sim$24 nm and $\sim$23 nm RMS. Qualitatively, the wavefront improvement can be observed in the more symmetric leakage and phase diversity PSFs, and the dark hole is better defined as aberrated structure at a few $\lambda / D$ has been removed. Quantitatively, the peak flux of the leakage increased by $\sim10 \%$, and the raw contrast at 2.5 $\lambda / D$ improved from $\sim$$2 \cdot 10^{-3}$ to $\sim$$6 \cdot 10^{-4}$ (solid lines in \autoref{fig:lab_results} c). Only the lowest Zernike modes (both even and odd) were corrected, which is clearly visible as the contrast improvement decreases with $\lambda / D$. The dashed-dotted lines in \autoref{fig:lab_results} c show the results of a similar correction performed at another moment in time, but here the images used for the raw contrast determination were saturated such that the raw contrast was speckle-noise limited, instead of read noise limited as with the solid lines. The bottom row shows the result of a closed-loop correction where we introduced and estimated a 300 nm RMS WFE with 50 Zernike modes, which should be correctable according to \autoref{fig:wfe_sens}. Images b and c show the pre- and post-correction after 5 iterations of closed loop correction. The total duration of these corrections was around 4 minutes. These figures show that the algorithm can correct large WFE, recovering a PSF that is qualitatively very similar to what is shown in \autoref{fig:lab_results} b. As shown in \autoref{fig:lab_results} f, we recover similar or deeper raw contrasts compared to the nominal system after correction, \autoref{fig:lab_results} c. In \autoref{fig:wavefront_aberrated_internal} b and c, the introduced and estimated wavefronts are shown. The WFE for these figures are respectively $\sim$293 nm and $\sim$311 nm. Note that the estimated wavefront also includes WFE already present in the system before adding the known aberration, as this was not corrected before the test. Tests conducted on another day suggest that the remaining WFE after correction is on the order of $\sim$59 nm RMS. These tests consist of sequential measurements of the nominal wavefront and a 150 nm RMS introduced wavefront error. These measurements were then compared to find a residual WFE. This error consists of evolved NCPA between measurements, DM calibration errors and algorithm errors. We did not investigate the amplitude of the individual error terms.

\begin{figure*}[ht]
\centering
   \includegraphics[width=17cm]{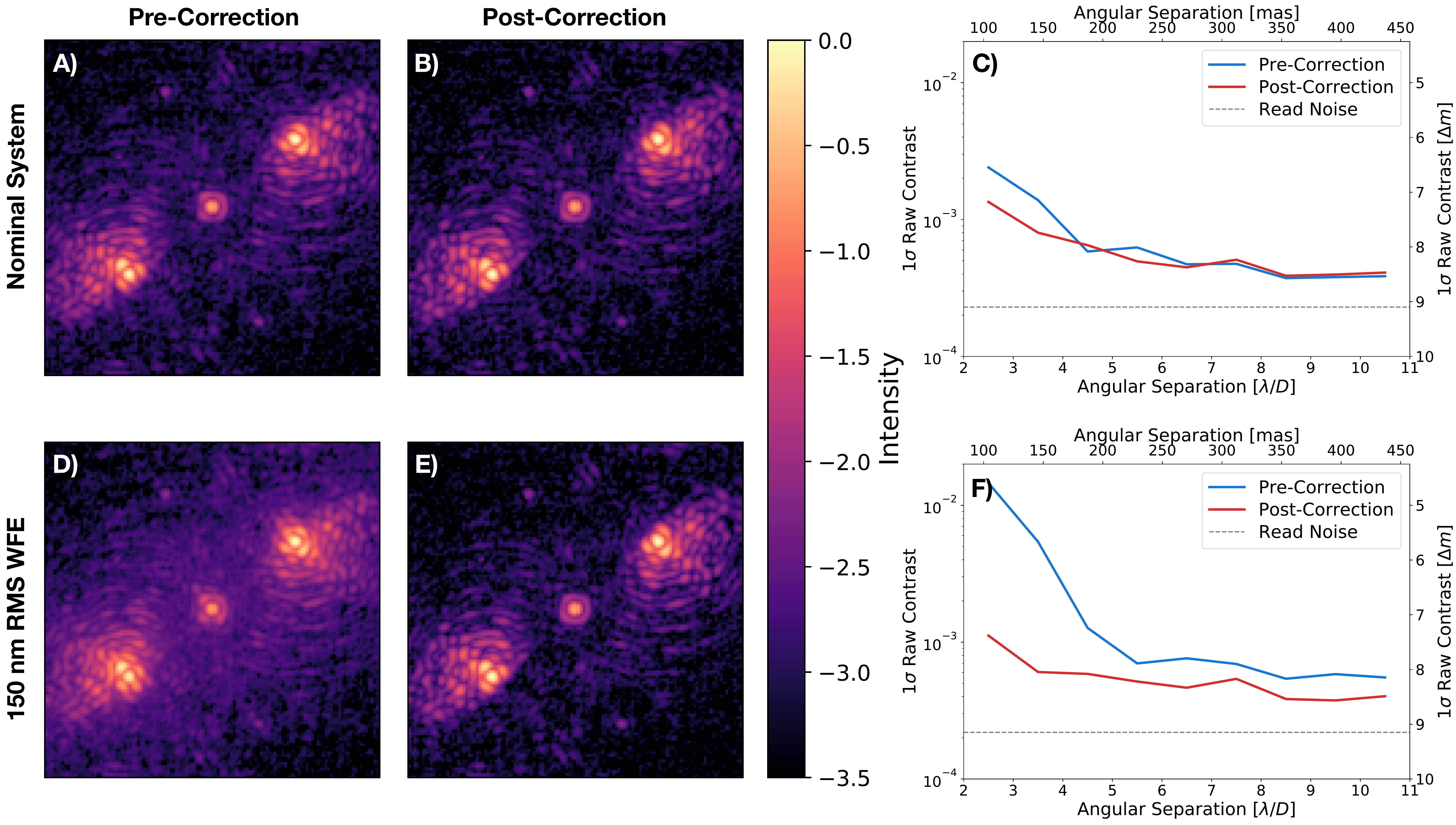}
     \caption{Focal-plane images and 1$\sigma$ raw contrast curves pre- and post-correction during on-sky observations. (a) Pre- and (b) post-correction images for the nominal SCExAO system. (c) The 1$\sigma$ raw contrast curves for the correction of the nominal system. The horizontal dashed line denotes the read noise level. (d) Pre- and (e) post-correction images for an additional wavefront error of 150 nm RMS. (f) The 1$\sigma$ raw contrast curves for the correction of the system with the additional 150 nm RMS wavefront error, with again the horizontal dashed line denoting the read noise. }
     \label{fig:onsky_results}
\end{figure*}

\begin{figure*}[ht]
\centering
   \includegraphics[width=17cm]{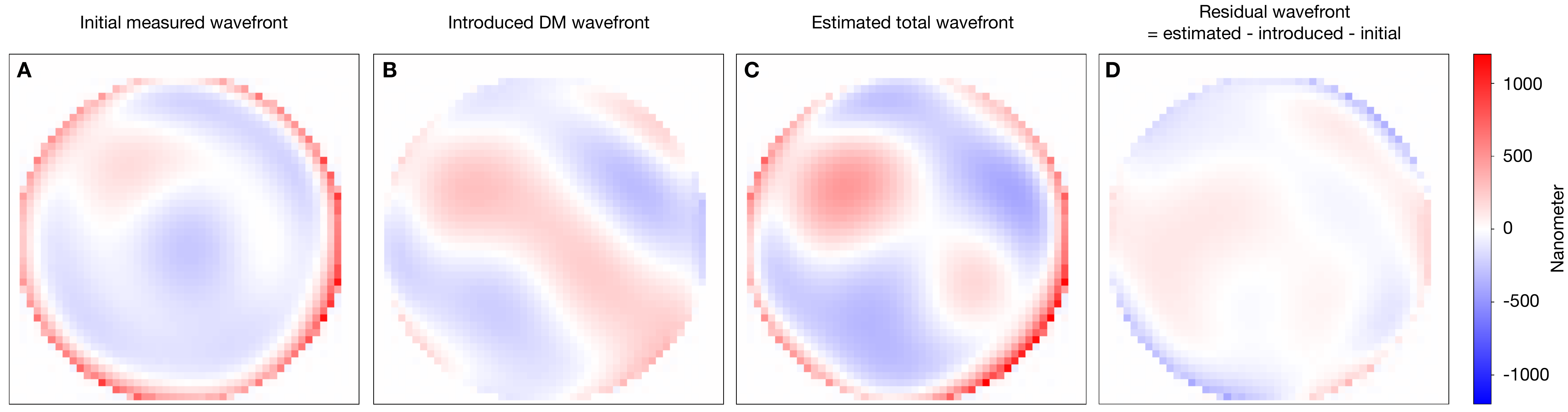}
     \caption{(a) Estimated wavefront of the nominal SCExAO system during on-sky observations and reconstructed by combining the estimated modal coefficients of the 30 Zernike modes. The estimated WFE is $\sim$187 nm RMS. The (b) introduced, (c) estimated and (d) residual wavefront from closed-loop tests during on-sky observations. (b) The introduced wavefront, consisting of 30 Zernike modes, with an RMS WFE of $\sim$146 nm. (c) The estimated wavefront, using the same 30 Zernike modes, with an RMS WFE of $\sim$244 nm. (d) The residual wavefront (introduced minus estimated and nominal wavefronts), the RMS WFE is $\sim$78 nm.}
     \label{fig:wavefront_aberrated_onsky}
\end{figure*}

\subsection{On-sky demonstration}\label{sec:onsky} 
On the 11th of January 2019 we observed Regulus ($m_H = 1.57$; \citealt{ducati2002vizier}) during a SCExAO engineering night to test the vAPP FPWFS on-sky, to evaluate the performance with an incoherent background due to residual turbulence and long exposure images, and to demonstrate with the true end-to-end optical system. Observing conditions were good - the seeing was $\sim0.2'' -0.4"$ - and predictive control was running \citep{guyon2017adaptive} during the tests. We conducted similar tests as with the internal source of which the results are shown in \autoref{fig:onsky_results}. The top row shows the results after 5 iterations of closed-loop with a gain=0.5 with the nominal SCExAO system when estimating 30 Zernike modes. The 5 iterations took approximately 4 minutes. Qualitatively, the first Airy ring of the leakage PSF becomes more rounded and symmetric, the coronagraphic PSFs are more similar, the edge of the dark hole becomes better defined and some speckles in the dark hole are removed. The gain in peak flux of the leakage PSF is 6\%. Shown in \autoref{fig:onsky_results} c, the contrast gain is moderate, improving by a factor $\sim$2 between 2.5 and 4 $\lambda/D$. The estimated wavefront is shown in \autoref{fig:wavefront_aberrated_onsky} a and the WFE is $\sim$187 nm RMS. A notable structure in this estimated wavefront is the sharp increase in phase at the edges of the pupil. It is thought that this phase ring is a relatively static and real structure and originates from the upstream AO188 system. This is because the PYWFS reference offset contains a similar structure as well. The bottom row shows the correction of low order quasi-static errors from the instrument on-sky plus and additional 150 nm RMS WFE that was added to the DM consisting of 30 Zernike modes. After 5 iterations (gain = 0.5), shown in \autoref{fig:onsky_results} e, the algorithm recovers a similar PSF in a similar time as for the correction of the nominal system (\autoref{fig:onsky_results} c), both in morphology and in achieved raw contrast, which is shown in \autoref{fig:onsky_results} f.  In \autoref{fig:wavefront_aberrated_onsky} b-d the introduced, estimated and residual wavefronts are shown, with the WFE being respectively $\sim$146 nm, $\sim$244 nm and $\sim$78 nm RMS. The residual wavefront was determined by subtracting the introduced and initial wavefronts from the estimated wavefront. This gives an estimate of the remaining wavefront error in the system. As discussed with the internal source results, the residual wavefront consists of evolved NCPA between measurements, DM calibration errors and algorithm errors. Note that the reported raw contrasts in \autoref{fig:onsky_results} c and f are worse than in \autoref{fig:lab_results} c and f due to the incoherent speckle background created by the uncorrected, residual atmospheric wavefront errors. 


\section{Discussion and Conclusion}\label{sec:conclusions} 
We have shown in \autoref{sec:theory} that an asymmetry in the pupil amplitude enables the vAPP coronagraphic PSFs to measure both even and odd pupil phase modes, generalizing the Asymmetric Pupil Fourier Wavefront Sensor (APF-WFS; \citealt{martinache2013asymmetric}) with spatial Linear Dark Field Control (LDFC; \citealt{miller2017spatial}) in the vAPP. The physical model for non-linear wavefront estimation developed in \autoref{sec:model} was tested in idealised simulations (\autoref{sec:simulations}), confirming that the vAPP currently installed at SCExAO can sense the even modes, but is more sensitive to odd modes. Simulations suggest that the algorithm should be able to reach one-shot correction resulting in a $ < \lambda/1000$ nm RMS WFE with approximately $10^7$ photons, and that with one iteration it should be able to correct up to $\sim\lambda/8$ of RMS WFE. In \autoref{sec:demonstration}, we have demonstrated the principle with the vAPP in SCExAO, both with the internal source and on-sky, by measuring and controlling the 30 lowest Zernike modes and improving the raw contrast between 2.5 and 4 $\lambda/D$ by a factor $\sim$2. Furthermore, we have shown that the algorithm can correct WFE's of 300 nm and 150 nm RMS with the internal source and on-sky, respectively. Although the contrast gains, both with the internal source and on-sky, are moderate, it does demonstrate that the coronagraphic PSFs of the vAPP can be used for wavefront sensing.\\
\indent The FPWFS performance of the SCExAO vAPP can be improved in various ways: 1) better model calibration to increase the contrast gain, 2) measuring more and higher spatial frequency modes to increase the area of correction, and 3) improving the convergence speed of the algorithm. For this demonstration the model was calibrated ad-hoc by tuning the pupil undersizing, rotation and the detector pixel scale by hand. These parameters should be more accurately measured or be part of the fitted parameters. To measure more and higher spatial frequency modes we want to replace the current Zernike mode basis with a pixel-mode basis \citep{paul2013high}. This should also reduce the effect of unsensed modes on the estimation, when estimating a truncated mode basis \citep{paul2013coronagraphic}. Improving the convergence speed of the algorithm would result in a higher correction cadence or the correction of a larger mode basis in a similar time. A relatively simple improvement would be using the multi-processing capabilities in the latest python libraries not yet available on ScexAO, this would enable multi-core processing and give a maximum factor of 40 speed improvement. Theoretically, the best possible performance we can then expect is a wavefront estimation with sub 1 nm WFE every six seconds for targets down to a $m_H=8$ (five seconds of integration and one second for estimation). Other coronagraph model improvements include the integral field spectrograph CHARIS, an accurate model of the focal-plane filter and the incoherent background due to residual turbulence effects. This is because science observations with the SCExAO vAPP are done with CHARIS. To prevent detector saturation due to bright fields of the vAPP, it requires a focal-plane neutral density filter that attenuates the bright fields. Therefore, exposures with the focal-plane filter in place will have a larger dynamic range. This increase in dynamic range will make the incoherent background, due to uncorrected atmospheric turbulence, a more prominent feature in the image compared to the current on-sky results. Thus, the incoherent background needs to be included within the coronagraph model as well \citep{herscovici2017analytic}, also including any asymmetries in this background \citep{cantalloube2018origin}. \\
\indent In this article we have focused on a FPWFS based on a grating-vAPP. We operated it only in narrowband mode or in an integral field spectrograph. Therefore, the FPWFS algorithm does not have to operate over broad wavelength ranges and the current coronagraph model would suffice. However, there are broadband imaging vAPPs implementations foreseen, as is detailed in \cite{bos2018fully}. These coronagraphs are more complicated in their optical design and therefore require a more advanced coronagraph model. For such broadband coronagraphs, the coronagraph model also needs to be extended to handle broadband FPWFS to take full advantage of the coronagraph. This can be done by evaluating the model at multiple wavelengths \citep{seldin2000closed} and would come at an increased computational cost. Operating in a broadband mode would increase the sensitivity as there is more light available, but speckles at larger $\lambda/D$ will be washed out and therefore higher frequency aberrations will be harder to measure. \\
\indent In this paper we discussed only FPWFS for vAPPs that generate asymmetric dark holes. But FPWFS with pupil-plane coronagraphs that create symmetric dark holes such as the $360^{\circ}$ vAPP \citep{otten2014vector} and the shaped pupil \citep{kasdin2007shaped} are desirable as well. Optimal designs of vAPPs with symmetric dark holes consist of only 0 and $\pi$ phase structures \citep{por2017optimal}. This results in a purely real pupil-plane electric field (\autoref{eq:pup_real_imag}) and thus if the aperture is symmetric the focal-plane electric field of such a vAPP will be completely imaginary (\autoref{tab:FraunhoferSymmetries}) and does not support a FPWFS. Therefore, similar to one-sided dark holes, FPWFSing is enabled when an amplitude asymmetry is introduced. A similar argument can be given for shaped pupil coronagraphs, as these  manipulate pupil-plane amplitude and thus only have a real pupil-plane electric field. A more important difference with one-sided dark hole PSFs is that such coronagraphs will not have a bright field in the other polarization that covers the dark hole of the coronagraph and thus does not directly probe the region of interest. \\
\indent Future work will investigate optimization of the aperture asymmetry to enhance the wavefront sensing capabilities of the coronagraph. This also has implications for the APF-WFS and spatial LDFC controlling dark holes dug with other methods than the vAPP. With the current implementation of the vAPP and algorithm we can only sense phase aberrations. Uncorrected amplitude errors due to the atmosphere and instrumental errors will limit the raw contrast to $\sim10^{-5}$ \citep{guyon2018extreme}. This is at a level that is not yet reached by the vAPP and therefore, sensing only phase aberrations is sufficient for now. When this raw contrast is met it would be necessary to sense both pupil amplitude and phase aberrations. It is expected that the two coronagraphic PSFs alone do not contain sufficient diversity to estimate both in one image. For the current SCExAO vAPP design, an additional classical phase diversity image will likely provide the required diversity to enable amplitude estimation \citep{herscovici2018experimental}, but will lower the science duty cycle. A better solution for future FPWFS vAPPs would be the addition of two strong phase diversity holograms, this would keep the science duty cycle at 100$\%$ at the cost of science throughput. Another use of FPWFS with the vAPP could be the fine co-phasing of multi-mirror telescopes from the image plane \citep{pope2014demonstration}, such as for the upcoming Giant Magellan Telescope. \\ 
\indent Integration of FPWFS with coronagraphy is a crucial step in the system wide integration of all optical modalities (e.g. spectroscopy and polarimetry) to get the best possible performance of high contrast imaging instruments. Major advantages are that NCPA can be measured up to the science focal-plane, and a 100$\%$ science duty cycle as science observations do not have to be interrupted to probe the dark hole. The vAPP has now shown to be able to combine both FPWFS and coronagraphy in one optic. 

\begin{acknowledgements} 
The authors thank the referee for comments on the manuscript that significantly improved the presentation of the work. The authors warmly thank S.Y. Haffert and E.H. Por for useful discussion on focal-plane wavefront sensing. The research of Steven P. Bos, David S. Doelman, and Frans Snik leading to these results has received funding from the European Research Council under ERC Starting Grant agreement 678194 (FALCONER). The development of SCExAO was supported by the JSPS (Grant-in-Aid for Research $\#$23340051, $\#$26220704, $\#$23103002), the Astrobiology Center (ABC) of the National Institutes of Natural Sciences, Japan, the Mt Cuba Foundation and the directors contingency fund at Subaru Telescope. The authors wish to recognize and acknowledge the very significant cultural role and reverence that the summit of Maunakea has always had within the indigenous Hawaiian community. We are most fortunate to have the opportunity to conduct observations from this mountain. This research made use of HCIPy, an open-source object-oriented framework written in Python for performing end-to-end simulations of high-contrast imaging instruments \citep{por2018hcipy}. This research used the following Python libaries: Scipy \citep{jones2014scipy}, Numpy \citep{walt2011numpy} and Matplotlib \citep{Hunter:2007}.
\end{acknowledgements}

\bibliography{report} 

\begin{thebibliography}{82}
\expandafter\ifx\csname natexlab\endcsname\relax\def\natexlab#1{#1}\fi

\bibitem[{Baudoz {et~al.}(2005)Baudoz, Boccaletti, Baudrand, \&
  Rouan}]{baudoz2005self}
Baudoz, P., Boccaletti, A., Baudrand, J., \& Rouan, D. 2005, Proceedings of the
  International Astronomical Union, 1, 553

\bibitem[{Berry(1987)}]{berry1987adiabatic}
Berry, M.~V. 1987, Journal of Modern Optics, 34, 1401

\bibitem[{Beuzit {et~al.}(2019)Beuzit, Vigan, Mouillet, Dohlen, Gratton,
  Boccaletti, Sauvage, Schmid, Langlois, Petit, {et~al.}}]{beuzit2019sphere}
Beuzit, J.-L., Vigan, A., Mouillet, D., {et~al.} 2019, arXiv preprint
  arXiv:1902.04080

\bibitem[{Boehle {et~al.}(2018)Boehle, Glauser, Kenworthy, Snik, Doelman,
  Quanz, \& Meyer}]{boehle2018cryogenic}
Boehle, A., Glauser, A.~M., Kenworthy, M.~A., {et~al.} 2018, in Ground-based
  and Airborne Instrumentation for Astronomy VII, Vol. 10702, International
  Society for Optics and Photonics, 107023Y

\bibitem[{Bord{\'e} \& Traub(2006)}]{borde2006high}
Bord{\'e}, P.~J. \& Traub, W.~A. 2006, The Astrophysical Journal, 638, 488

\bibitem[{Bos {et~al.}(2018)Bos, Doelman, de~Boer, Por, Norris, Escuti, \&
  Snik}]{bos2018fully}
Bos, S.~P., Doelman, D.~S., de~Boer, J., {et~al.} 2018, in Advances in Optical
  and Mechanical Technologies for Telescopes and Instrumentation III, Vol.
  10706, International Society for Optics and Photonics, 107065M

\bibitem[{Byrd {et~al.}(1995)Byrd, Lu, Nocedal, \& Zhu}]{byrd1995limited}
Byrd, R.~H., Lu, P., Nocedal, J., \& Zhu, C. 1995, SIAM Journal on Scientific
  Computing, 16, 1190

\bibitem[{Cantalloube {et~al.}(2018)Cantalloube, Por, Dohlen, Sauvage, Vigan,
  Kasper, Bharmal, Henning, Brandner, Milli, {et~al.}}]{cantalloube2018origin}
Cantalloube, F., Por, E., Dohlen, K., {et~al.} 2018, Astronomy \& Astrophysics,
  620, L10

\bibitem[{Close {et~al.}(2018)Close, Males, Durney, Sauve, Kautz, Hedglen,
  Schatz, Lumbres, Miller, Van~Gorkom, {et~al.}}]{close2018optical}
Close, L.~M., Males, J.~R., Durney, O., {et~al.} 2018, arXiv preprint
  arXiv:1807.04311

\bibitem[{Codona \& Doble(2012)}]{codona2012experimental}
Codona, J.~L. \& Doble, N. 2012, in Adaptive Optics Systems III, Vol. 8447,
  International Society for Optics and Photonics, 84476R

\bibitem[{C{\^o}t{\'e} {et~al.}(2018)C{\^o}t{\'e}, Allain, Brousseau, Lord,
  Ouahbi, Ouellet, Patel, Thibault, Vall{\'e}e, Belikov,
  {et~al.}}]{cote2018precursor}
C{\^o}t{\'e}, O., Allain, G., Brousseau, D., {et~al.} 2018, in Ground-based and
  Airborne Instrumentation for Astronomy VII, Vol. 10702, International Society
  for Optics and Photonics, 1070248

\bibitem[{Davies {et~al.}(2018)Davies, Alves, Cl{\'e}net, Lang-Bardl, Nicklas,
  Pott, Ragazzoni, Tolstoy, Amico, Anwand-Heerwart,
  {et~al.}}]{davies2018micado}
Davies, R., Alves, J., Cl{\'e}net, Y., {et~al.} 2018, in Ground-based and
  Airborne Instrumentation for Astronomy VII, Vol. 10702, International Society
  for Optics and Photonics, 107021S

\bibitem[{De~Kok {et~al.}(2011)De~Kok, Stam, \&
  Karalidi}]{de2011characterizing}
De~Kok, R., Stam, D., \& Karalidi, T. 2011, The Astrophysical Journal, 741, 59

\bibitem[{Doelman {et~al.}(2019)Doelman, Auer, Escuti, \&
  Snik}]{doelman2019simultaneous}
Doelman, D.~S., Auer, F.~F., Escuti, M.~J., \& Snik, F. 2019, Optics letters,
  44, 17

\bibitem[{Doelman {et~al.}(2017)Doelman, Snik, Warriner, \&
  Escuti}]{doelman2017patterned}
Doelman, D.~S., Snik, F., Warriner, N.~Z., \& Escuti, M.~J. 2017, in Techniques
  and Instrumentation for Detection of Exoplanets VIII, Vol. 10400,
  International Society for Optics and Photonics, 104000U

\bibitem[{Ducati(2002)}]{ducati2002vizier}
Ducati, J. 2002, VizieR Online Data Catalog, 2237

\bibitem[{Feautrier {et~al.}(2017)Feautrier, Gach, Greffe, Clop, Lemarchand,
  Carmignani, Stadler, Doucour{\'e}, \& Boutolleau}]{feautrier2017c}
Feautrier, P., Gach, J.-L., Greffe, T., {et~al.} 2017, in Image Sensing
  Technologies: Materials, Devices, Systems, and Applications IV, Vol. 10209,
  International Society for Optics and Photonics, 102090G

\bibitem[{Galicher {et~al.}(2008)Galicher, Baudoz, \&
  Rousset}]{galicher2008wavefront}
Galicher, R., Baudoz, P., \& Rousset, G. 2008, Astronomy \& Astrophysics, 488,
  L9

\bibitem[{Gonsalves(1982)}]{gonsalves1982phase}
Gonsalves, R.~A. 1982, Optical Engineering, 21, 215829

\bibitem[{Goodman(2005)}]{goodman2005introduction}
Goodman, J.~W. 2005, Introduction to Fourier optics (Roberts and Company
  Publishers)

\bibitem[{Groff {et~al.}(2014)Groff, Kasdin, Limbach, Galvin, Carr, Knapp,
  Brandt, Loomis, Jarosik, Mede, {et~al.}}]{groff2014construction}
Groff, T.~D., Kasdin, N.~J., Limbach, M.~A., {et~al.} 2014, in Ground-based and
  Airborne Instrumentation for Astronomy V, Vol. 9147, International Society
  for Optics and Photonics, 91471W

\bibitem[{Groff {et~al.}(2015)Groff, Riggs, Kern, \& Kasdin}]{groff2015methods}
Groff, T.~D., Riggs, A.~E., Kern, B., \& Kasdin, N.~J. 2015, Journal of
  Astronomical Telescopes, Instruments, and Systems, 2, 011009

\bibitem[{Guyon(2005)}]{guyon2005limits}
Guyon, O. 2005, The Astrophysical Journal, 629, 592

\bibitem[{Guyon(2018)}]{guyon2018extreme}
Guyon, O. 2018, Annual Review of Astronomy and Astrophysics, 56, 315

\bibitem[{Guyon \& Males(2017)}]{guyon2017adaptive}
Guyon, O. \& Males, J. 2017, arXiv preprint arXiv:1707.00570

\bibitem[{Guyon {et~al.}(2017)Guyon, Miller, Males, Belikov, \&
  Kern}]{guyon2017spectral}
Guyon, O., Miller, K., Males, J., Belikov, R., \& Kern, B. 2017, arXiv preprint
  arXiv:1706.07377

\bibitem[{Guyon {et~al.}(2018)Guyon, Sevin, Ltaief, Skaf, Martinache,
  Gratadour, Cetre, Males, Lozi, Clergeon, {et~al.}}]{guyon2018compute}
Guyon, O., Sevin, A., Ltaief, H., {et~al.} 2018, in Adaptive Optics Systems VI,
  Vol. 10703, International Society for Optics and Photonics, 107031E

\bibitem[{Haffert {et~al.}(2019)Haffert, Bohn, de~Boer, Snellen, Brinchmann,
  Girard, Keller, \& Bacon}]{haffert2019two}
Haffert, S., Bohn, A., de~Boer, J., {et~al.} 2019, Nature Astronomy, 1

\bibitem[{Haffert {et~al.}(2018)Haffert, Wilby, Keller, Snellen, Doelman, Por,
  van Kooten, Bos, \& Wardenier}]{haffert2018sky}
Haffert, S., Wilby, M., Keller, C., {et~al.} 2018, in Adaptive Optics Systems
  VI, Vol. 10703, International Society for Optics and Photonics, 1070323

\bibitem[{Herscovici-Schiller {et~al.}(2018)Herscovici-Schiller, Mugnier,
  Baudoz, Galicher, Sauvage, \& Paul}]{herscovici2018experimental}
Herscovici-Schiller, O., Mugnier, L.~M., Baudoz, P., {et~al.} 2018, Astronomy
  \& Astrophysics, 614, A142

\bibitem[{Herscovici-Schiller {et~al.}(2017)Herscovici-Schiller, Mugnier, \&
  Sauvage}]{herscovici2017analytic}
Herscovici-Schiller, O., Mugnier, L.~M., \& Sauvage, J.-F. 2017, Monthly
  Notices of the Royal Astronomical Society: Letters, 467, L105

\bibitem[{Hoeijmakers {et~al.}(2018)Hoeijmakers, Schwarz, Snellen, de~Kok,
  Bonnefoy, Chauvin, Lagrange, \& Girard}]{hoeijmakers2018medium}
Hoeijmakers, H., Schwarz, H., Snellen, I., {et~al.} 2018, Astronomy \&
  Astrophysics, 617, A144

\bibitem[{Hunter(2007)}]{Hunter:2007}
Hunter, J.~D. 2007, Computing In Science \& Engineering, 9, 90

\bibitem[{Jones {et~al.}(2014)Jones, Oliphant, \& Peterson}]{jones2014scipy}
Jones, E., Oliphant, T., \& Peterson, P. 2014

\bibitem[{Jovanovic {et~al.}(2018)Jovanovic, Absil, Baudoz, Beaulieu, Bottom,
  Cady, Carlomagno, Carlotti, Doelman, Fogarty, Galicher, Guyon, Haffert, Huby,
  Jewell, Keller, A.~Kenworthy, Knight, Kuhn, \& Ygouf}]{Jovanovic2018}
Jovanovic, N., Absil, O., Baudoz, P., {et~al.} 2018, in Proc. {{SPIE}}, Vol.
  10703, Adaptive Optics Systems VI

\bibitem[{Jovanovic {et~al.}(2015)Jovanovic, Martinache, Guyon, Clergeon,
  Singh, Kudo, Garrel, Newman, Doughty, Lozi, {et~al.}}]{jovanovic2015subaru}
Jovanovic, N., Martinache, F., Guyon, O., {et~al.} 2015, Publications of the
  Astronomical Society of the Pacific, 127, 890

\bibitem[{Kanseri {et~al.}(2008)Kanseri, Bisht, Kandpal, \&
  Rath}]{kanseri2008observation}
Kanseri, B., Bisht, N.~S., Kandpal, H., \& Rath, S. 2008, American Journal of
  Physics, 76, 39

\bibitem[{Kasdin {et~al.}(2007)Kasdin, Vanderbei, \&
  Belikov}]{kasdin2007shaped}
Kasdin, N.~J., Vanderbei, R.~J., \& Belikov, R. 2007, Comptes Rendus Physique,
  8, 312

\bibitem[{Kenworthy {et~al.}(2018)Kenworthy, Absil, Carlomagno, Ag{\'o}cs, Por,
  Bos, Brandl, \& Snik}]{kenworthy2018review}
Kenworthy, M.~A., Absil, O., Carlomagno, B., {et~al.} 2018, in Ground-based and
  Airborne Instrumentation for Astronomy VII, Vol. 10702, International Society
  for Optics and Photonics, 10702A3

\bibitem[{Lozi {et~al.}(2018)Lozi, Guyon, \& et~al.}]{lozi2018scexao}
Lozi, J., Guyon, O., \& et~al. 2018, in Adaptive Optics Systems VI, Vol. 10703,
  International Society for Optics and Photonics

\bibitem[{Macintosh {et~al.}(2014)Macintosh, Graham, Ingraham, Konopacky,
  Marois, Perrin, Poyneer, Bauman, Barman, Burrows,
  {et~al.}}]{macintosh2014first}
Macintosh, B., Graham, J.~R., Ingraham, P., {et~al.} 2014, Proceedings of the
  National Academy of Sciences, 111, 12661

\bibitem[{Males {et~al.}(2018)Males, Close, Miller, Schatz, Doelman, Lumbres,
  Snik, Rodack, Knight, Van~Gorkom, {et~al.}}]{males2018magao}
Males, J.~R., Close, L.~M., Miller, K., {et~al.} 2018, in Adaptive Optics
  Systems VI, Vol. 10703, International Society for Optics and Photonics,
  1070309

\bibitem[{Marois {et~al.}(2006)Marois, Lafreniere, Doyon, Macintosh, \&
  Nadeau}]{marois2006angular}
Marois, C., Lafreniere, D., Doyon, R., Macintosh, B., \& Nadeau, D. 2006, The
  Astrophysical Journal, 641, 556

\bibitem[{Martinache(2013)}]{martinache2013asymmetric}
Martinache, F. 2013, Publications of the Astronomical Society of the Pacific,
  125, 422

\bibitem[{Martinache {et~al.}(2016)Martinache, Jovanovic, \&
  Guyon}]{martinache2016closed}
Martinache, F., Jovanovic, N., \& Guyon, O. 2016, Astronomy \& Astrophysics,
  593, A33

\bibitem[{Martinez {et~al.}(2013)Martinez, Kasper, Costille, Sauvage, Dohlen,
  Puget, \& Beuzit}]{martinez2013speckle}
Martinez, P., Kasper, M., Costille, A., {et~al.} 2013, Astronomy \&
  Astrophysics, 554, A41

\bibitem[{Martinez {et~al.}(2012)Martinez, Loose, Carpentier, \&
  Kasper}]{martinez2012speckle}
Martinez, P., Loose, C., Carpentier, E.~A., \& Kasper, M. 2012, Astronomy \&
  Astrophysics, 541, A136

\bibitem[{Mazoyer {et~al.}(2013)Mazoyer, Baudoz, Galicher, Mas, \&
  Rousset}]{mazoyer2013estimation}
Mazoyer, J., Baudoz, P., Galicher, R., Mas, M., \& Rousset, G. 2013, Astronomy
  \& Astrophysics, 557, A9

\bibitem[{Miller {et~al.}(2017)Miller, Guyon, \& Males}]{miller2017spatial}
Miller, K., Guyon, O., \& Males, J. 2017, Journal of Astronomical Telescopes,
  Instruments, and Systems, 3, 049002

\bibitem[{Miller {et~al.}(2018)Miller, Males, Guyon, Close, Doelman, Snik, Por,
  Wilby, Bohlman, Lumbres, {et~al.}}]{miller2018focal}
Miller, K., Males, J.~R., Guyon, O., {et~al.} 2018, in Adaptive Optics Systems
  VI, Vol. 10703, International Society for Optics and Photonics, 107031T

\bibitem[{Miller(2018)}]{miller2018development}
Miller, K.~L. 2018, PhD thesis, The University of Arizona

\bibitem[{Minowa {et~al.}(2010)Minowa, Hayano, Oya, Watanabe, Hattori, Guyon,
  Egner, Saito, Ito, Takami, {et~al.}}]{minowa2010performance}
Minowa, Y., Hayano, Y., Oya, S., {et~al.} 2010, in Adaptive Optics Systems II,
  Vol. 7736, International Society for Optics and Photonics, 77363N

\bibitem[{Mujat {et~al.}(2004)Mujat, Dogariu, \& Wolf}]{mujat2004law}
Mujat, M., Dogariu, A., \& Wolf, E. 2004, JOSA A, 21, 2414

\bibitem[{N'Diaye {et~al.}(2013)N'Diaye, Dohlen, Fusco, \&
  Paul}]{n2013calibration}
N'Diaye, M., Dohlen, K., Fusco, T., \& Paul, B. 2013, Astronomy \&
  Astrophysics, 555, A94

\bibitem[{N'Diaye {et~al.}(2018)N'Diaye, Martinache, Jovanovic, Lozi, Guyon,
  Norris, Ceau, \& Mary}]{n2018calibration}
N'Diaye, M., Martinache, F., Jovanovic, N., {et~al.} 2018, Astronomy \&
  Astrophysics, 610, A18

\bibitem[{N'Diaye {et~al.}(2016)N'Diaye, Vigan, Dohlen, Sauvage, Caillat,
  Costille, Girard, Beuzit, Fusco, Blanchard, {et~al.}}]{n2016zelda}
N'Diaye, M., Vigan, A., Dohlen, K., {et~al.} 2016, in Adaptive Optics Systems
  V, Vol. 9909, International Society for Optics and Photonics, 99096S

\bibitem[{Oh \& Escuti(2008)}]{oh2008achromatic}
Oh, C. \& Escuti, M.~J. 2008, Optics letters, 33, 2287

\bibitem[{Otten {et~al.}(2017)Otten, Snik, Kenworthy, Keller, Males, Morzinski,
  Close, Codona, Hinz, Hornburg, {et~al.}}]{otten2017sky}
Otten, G.~P., Snik, F., Kenworthy, M.~A., {et~al.} 2017, The Astrophysical
  Journal, 834, 175

\bibitem[{Otten {et~al.}(2014)Otten, Snik, Kenworthy, Miskiewicz, Escuti, \&
  Codona}]{otten2014vector}
Otten, G.~P., Snik, F., Kenworthy, M.~A., {et~al.} 2014, in Advances in Optical
  and Mechanical Technologies for Telescopes and Instrumentation, Vol. 9151,
  International Society for Optics and Photonics, 91511R

\bibitem[{Pancharatnam(1956)}]{pancharatnam1956generalized}
Pancharatnam, S. 1956, in Proceedings of the Indian Academy of Sciences-Section
  A, Vol.~44, Springer, 398--417

\bibitem[{Paul {et~al.}(2013{\natexlab{a}})Paul, Mugnier, Sauvage, Dohlen, \&
  Ferrari}]{paul2013high}
Paul, B., Mugnier, L., Sauvage, J.-F., Dohlen, K., \& Ferrari, M.
  2013{\natexlab{a}}, Optics Express, 21, 31751

\bibitem[{Paul {et~al.}(2013{\natexlab{b}})Paul, Sauvage, \&
  Mugnier}]{paul2013coronagraphic}
Paul, B., Sauvage, J.-F., \& Mugnier, L. 2013{\natexlab{b}}, Astronomy \&
  Astrophysics, 552, A48

\bibitem[{Paul {et~al.}(2014)Paul, Sauvage, Mugnier, Dohlen, Petit, Fusco,
  Mouillet, Beuzit, \& Ferrari}]{paul2014compensation}
Paul, B., Sauvage, J.-F., Mugnier, L., {et~al.} 2014, Astronomy \&
  Astrophysics, 572, A32

\bibitem[{Paxman {et~al.}(1992)Paxman, Schulz, \& Fienup}]{paxman1992joint}
Paxman, R.~G., Schulz, T.~J., \& Fienup, J.~R. 1992, JOSA A, 9, 1072

\bibitem[{Peters-Limbach {et~al.}(2013)Peters-Limbach, Groff, Kasdin, Driscoll,
  Galvin, Foster, Carr, LeClerc, Fagan, McElwain, {et~al.}}]{peters2013optical}
Peters-Limbach, M.~A., Groff, T.~D., Kasdin, N.~J., {et~al.} 2013, in
  Techniques and Instrumentation for Detection of Exoplanets VI, Vol. 8864,
  International Society for Optics and Photonics, 88641N

\bibitem[{Pope {et~al.}(2014)Pope, Cvetojevic, Cheetham, Martinache, Norris, \&
  Tuthill}]{pope2014demonstration}
Pope, B., Cvetojevic, N., Cheetham, A., {et~al.} 2014, Monthly Notices of the
  Royal Astronomical Society, 440, 125

\bibitem[{Por(2017)}]{por2017optimal}
Por, E.~H. 2017, in Techniques and Instrumentation for Detection of Exoplanets
  VIII, Vol. 10400, International Society for Optics and Photonics, 104000V

\bibitem[{Por {et~al.}(2018)Por, Haffert, Radhakrishnan, Doelman, Van~Kooten,
  \& Bos}]{por2018hcipy}
Por, E.~H., Haffert, S.~Y., Radhakrishnan, V.~M., {et~al.} 2018, in Proc.
  {{SPIE}}, Vol. 10703, Adaptive Optics Systems VI

\bibitem[{Por \& Keller(2016)}]{por2016focal}
Por, E.~H. \& Keller, C.~U. 2016, in Adaptive Optics Systems V, Vol. 9909,
  International Society for Optics and Photonics, 990959

\bibitem[{Ruane {et~al.}(2019)Ruane, Ngo, Mawet, Absil, Choquet, Cook,
  Gonzalez, Huby, Matthews, Meshkat, {et~al.}}]{ruane2019reference}
Ruane, G., Ngo, H., Mawet, D., {et~al.} 2019, The Astronomical Journal, 157,
  118

\bibitem[{Sauvage {et~al.}(2012)Sauvage, Mugnier, Paul, \&
  Villecroze}]{sauvage2012coronagraphic}
Sauvage, J.-F., Mugnier, L., Paul, B., \& Villecroze, R. 2012, Optics Letters,
  37, 4808

\bibitem[{Seldin {et~al.}(2000)Seldin, Paxman, Zarifis, Benson, \&
  Stone}]{seldin2000closed}
Seldin, J.~H., Paxman, R.~G., Zarifis, V.~G., Benson, L., \& Stone, R.~E. 2000,
  in Imaging technology and telescopes, Vol. 4091, International Society for
  Optics and Photonics, 48--63

\bibitem[{Snik \& Keller(2013)}]{snik2013astronomical}
Snik, F. \& Keller, C.~U. 2013, in Planets, Stars and Stellar Systems
  (Springer), 175--221

\bibitem[{Snik {et~al.}(2012)Snik, Otten, Kenworthy, Miskiewicz, Escuti,
  Packham, \& Codona}]{snik2012vector}
Snik, F., Otten, G., Kenworthy, M., {et~al.} 2012, in Modern Technologies in
  Space-and Ground-based Telescopes and Instrumentation II, Vol. 8450,
  International Society for Optics and Photonics, 84500M

\bibitem[{Sparks \& Ford(2002)}]{sparks2002imaging}
Sparks, W.~B. \& Ford, H.~C. 2002, The Astrophysical Journal, 578, 543

\bibitem[{Stam {et~al.}(2004)Stam, Hovenier, \& Waters}]{stam2004using}
Stam, D., Hovenier, J., \& Waters, L. 2004, Astronomy \& Astrophysics, 428, 663

\bibitem[{Traub \& Oppenheimer(2010)}]{traub2010direct}
Traub, W.~A. \& Oppenheimer, B.~R. 2010, Exoplanets, 111

\bibitem[{van Holstein {et~al.}(2017)van Holstein, Snik, Girard, de~Boer,
  Ginski, Keller, Stam, Beuzit, Mouillet, Kasper, {et~al.}}]{van2017combining}
van Holstein, R.~G., Snik, F., Girard, J.~H., {et~al.} 2017, in Techniques and
  Instrumentation for Detection of Exoplanets VIII, Vol. 10400, International
  Society for Optics and Photonics, 1040015

\bibitem[{Vigan {et~al.}(2015)Vigan, Gry, Salter, Mesa, Homeier, Moutou, \&
  Allard}]{vigan2015high}
Vigan, A., Gry, C., Salter, G., {et~al.} 2015, Monthly Notices of the Royal
  Astronomical Society, 454, 129

\bibitem[{Vigan {et~al.}(2018)Vigan, N'Diaye, Dohlen, Milli, Wahhaj, Sauvage,
  Beuzit, Pourcelot, Mouillet, \& Zins}]{vigan2018sky}
Vigan, A., N'Diaye, M., Dohlen, K., {et~al.} 2018, in Adaptive Optics Systems
  VI, Vol. 10703, International Society for Optics and Photonics, 107035O

\bibitem[{Walt {et~al.}(2011)Walt, Colbert, \& Varoquaux}]{walt2011numpy}
Walt, S. v.~d., Colbert, S.~C., \& Varoquaux, G. 2011, Computing in Science \&
  Engineering, 13, 22

\bibitem[{Wilby {et~al.}(2017)Wilby, Keller, Snik, Korkiakoski, \&
  Pietrow}]{wilby2017coronagraphic}
Wilby, M.~J., Keller, C.~U., Snik, F., Korkiakoski, V., \& Pietrow, A.~G. 2017,
  Astronomy \& Astrophysics, 597, A112

\end{thebibliography}
\bibliographystyle{aa} 

\begin{appendix}

\section{Phase retrieval examples }\label{sec:phaseretrieval}
\begin{figure*}
\centering
   \includegraphics[width=17cm]{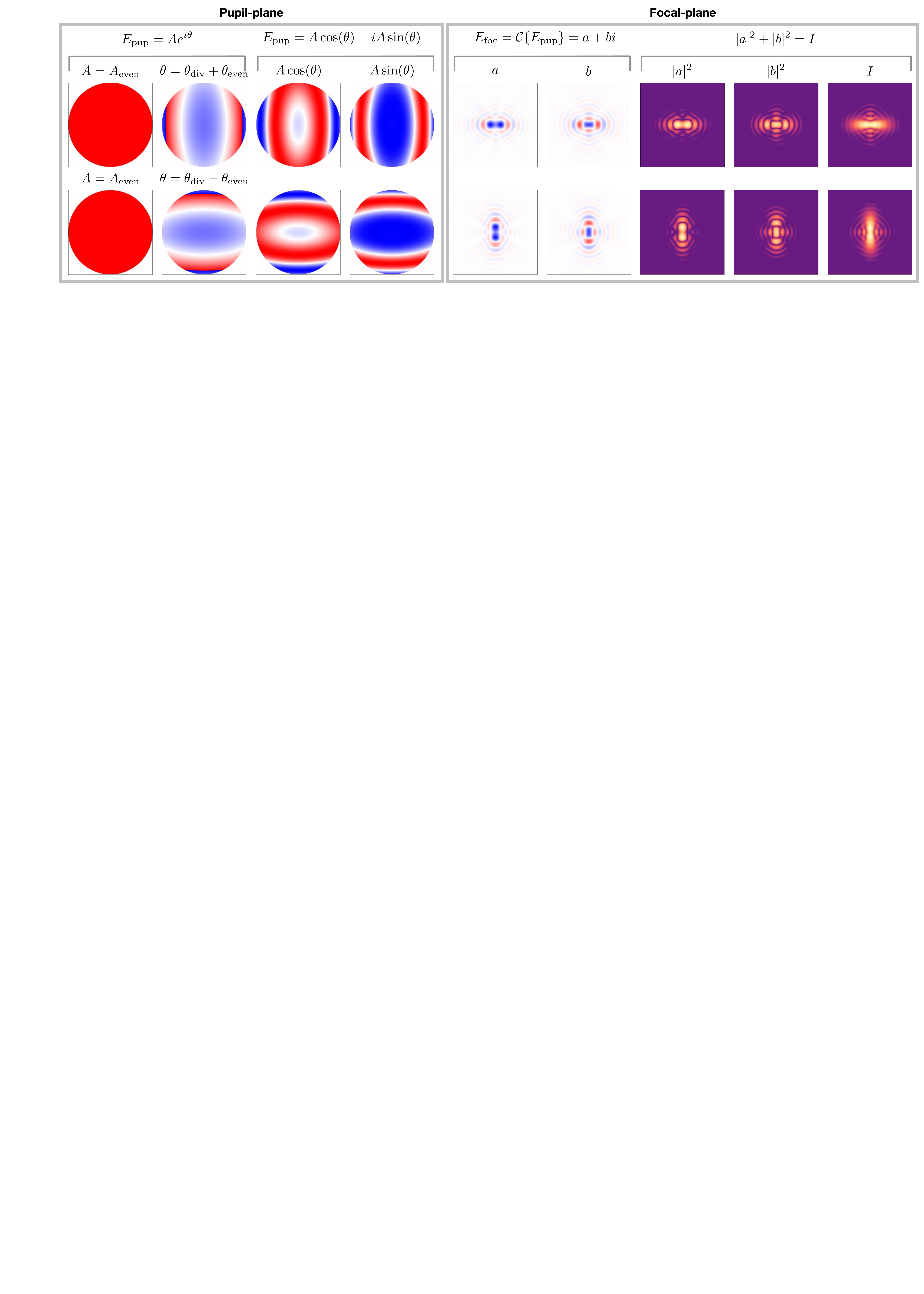}
     \caption{Focal- and pupil-plane quantities for a static diversity phase (defocus) with an even phase aberration (astigmatism) that has an alternating sign between the rows. The columns in the pupil-plane box show (from left to right) the amplitude, phase, real and imaginary electric field. In the focal-plane box, the columns show the real and imaginary electric fields, the power in the real and imaginary electric fields, and the total power.}
     \label{fig:theory_fig_D}
\end{figure*}
\begin{figure*}
\centering
   \includegraphics[width=17cm]{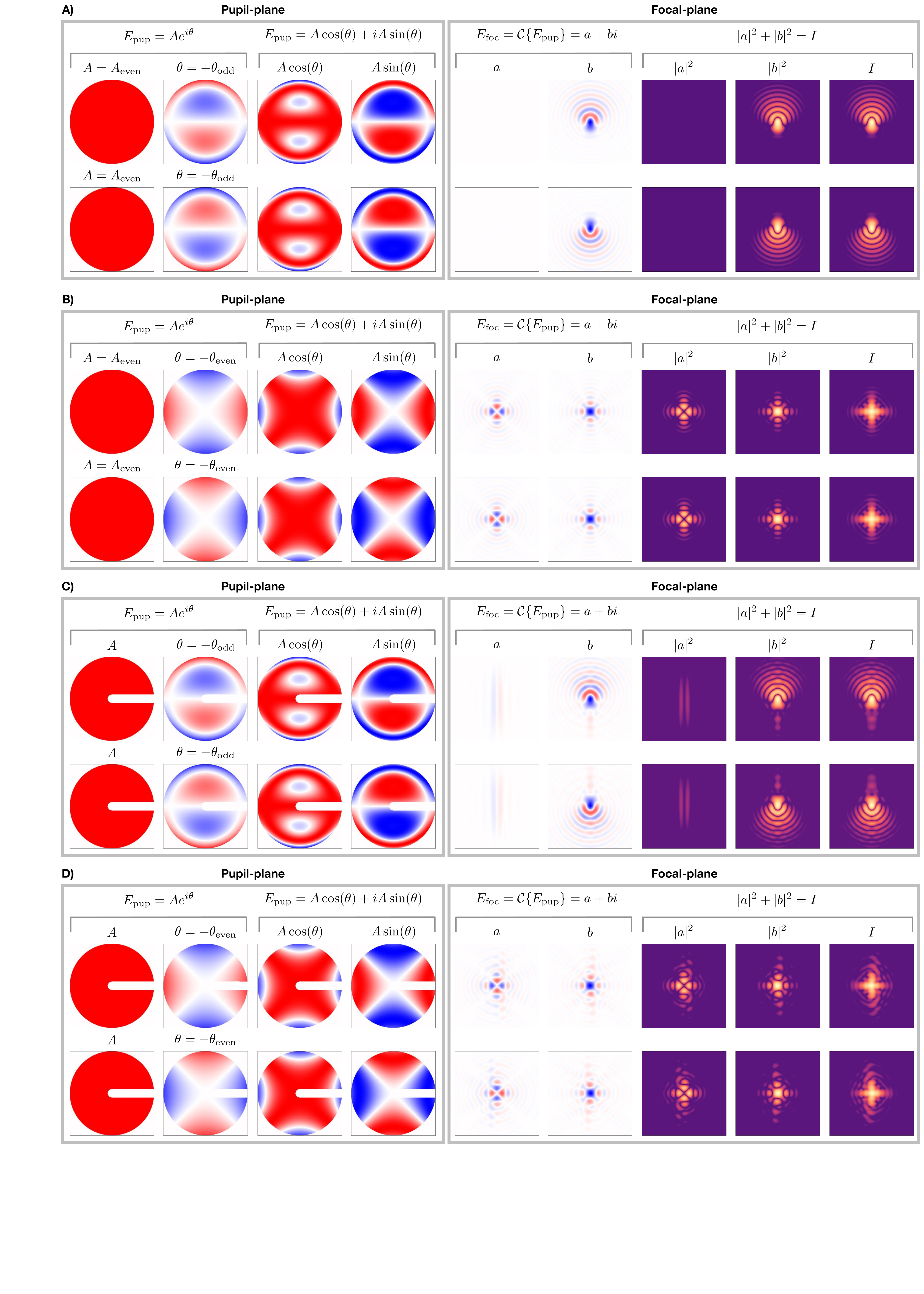}
     \caption{Focal- and pupil-plane quantities for different combinations of pupil symmetries and phase aberrations with alternating sign. (a) Symmetric aperture with odd aberration coma. (b) Symmetric aperture with even aberration astigmatism. (c) Asymmetric aperture with odd aberration coma. (d) Asymmetric aperture with even aberration astigmatism. The columns in the pupil-plane box show (from left to right) the amplitude, phase, real and imaginary electric field. In the focal-plane box, the columns show the real and imaginary electric fields, the power in the real and imaginary electric fields, and the total power.}
     \label{fig:theory_fig_C}
\end{figure*}
\begin{figure*}
\centering
   \includegraphics[width=17cm]{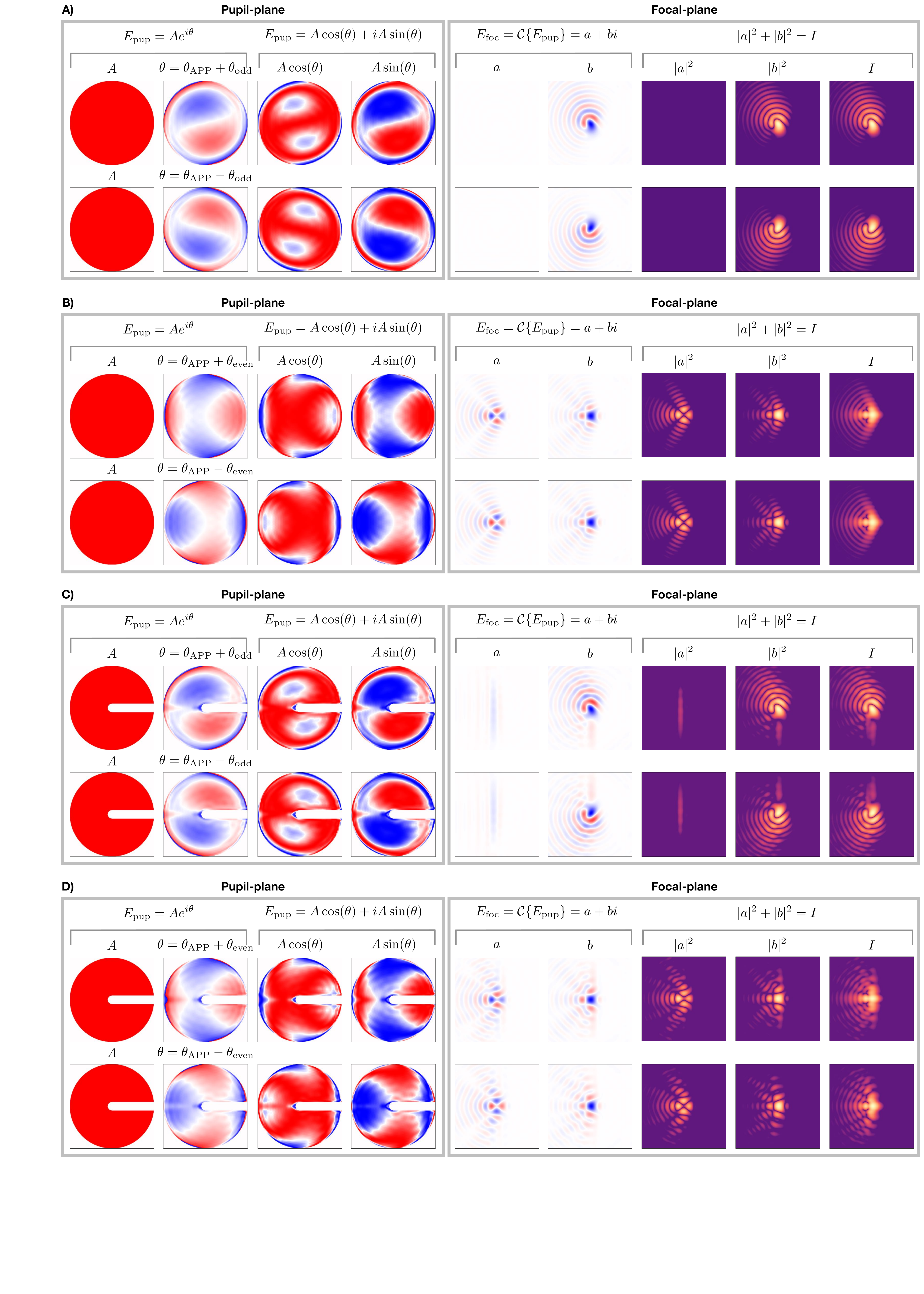}
     \caption{Focal- and pupil-plane quantities for APPs designed for different pupil symmetries subjected to phase aberrations with alternating sign. (a) Symmetric aperture with odd aberration coma. (b) Symmetric aperture with even aberration astigmatism. (c) Asymmetric aperture with odd aberration coma. (d) Asymmetric aperture with even aberration astigmatism. The columns in the pupil-plane box show (from left to right) the amplitude, phase, real and imaginary electric field. In the focal-plane box, the columns show the real and imaginary electric fields, the power in the real and imaginary electric fields, and the total power.}
     \label{fig:theory_fig_E}
\end{figure*}
In \autoref{sec:theory} we explained how the sign of even phase aberrations can be retrieved using an asymmetric pupil amplitude. In this appendix we will give a more complete set of examples of how pupil symmetries, and APPs designed for them, respond to even and odd aberrations. \\
\indent In the top row of \autoref{fig:theory_fig_C} a we visually present \Crefrange{eq:E_pup_tot}{eq:intensity_real_imag} for an even aperture (${A} = {A}_{\text{even}}$) and the odd coma aberration (${\theta} = {\theta}_{\text{odd}}$). The real part of the pupil-plane electric field, ${A} \cos({\theta})$, is completely even and contains no sign information due do $\cos(-\theta)$=$\cos(\theta)$; the imaginary part,  $i {A} \sin({\theta})$, is completely odd. Therefore, propagation to the focal-plane will result in a fully imaginary electric field $i{b}$ (see \autoref{tab:FraunhoferSymmetries}) that has even and odd components. The bottom row of \autoref{fig:theory_fig_C} a shows the same images but for a flipped sign of the phase (${\theta} = -{\theta}_{\text{odd}}$), equivalent to conjugating the pupil-plane electric field. Thus the phase conjugation results in a flipped PSF, and due to the odd component of ${E}_{\text{foc}}$, it results in a morphological change that can be measured, which is clearly visible in \autoref{fig:theory_fig_C} a. \\ 
\indent \autoref{fig:theory_fig_C} b shows similar figures, but for the even astigmatism aberration (${\theta}$=${\theta}_{\text{even}}$) and again, the top and bottom rows show opposite signs of the aberration. As expected, both real and imaginary parts of the pupil-plane electric field, ${A}\cos(\theta)$ and $i {A} \sin(\theta)$, are even. Again, the sign information is encoded in the imaginary pupil electric field term $i {A} \sin(\theta)$. This term propagates, in contrast to the odd phase aberration, to the real part of the focal-plane electric field ${a}$. Therefore, all sign information of the even phase aberration in encoded in the real focal-plane electric field. Because the aberration is even, this does not result in morphological changes in the total intensity measurement ${I}_{\text{foc}}$ and thus sign information of the even aberration cannot be measured. \\   
\indent This well-known sign ambiguity (\citealt{gonsalves1982phase};  \citealt{paxman1992joint}) is often broken by introducing a static, known diversity phase that results in a known real focal-plane electric field. This real electric field can interfere with the sign information carrying real electric field resulting from the even phase aberrations, which enables unique intensity changes for sign changes. \autoref{fig:theory_fig_D} shows the case of defocus as a static phase diversity and the even astigmatism aberration used in \autoref{fig:theory_fig_D}. It shows a significant morphology change in both the real and imaginary focal-plane power $|{a}|^2$ and $|{b}|^2$ that can be used to determine sign changes. \\
\indent In \autoref{sec:theory} we explained how the Asymmetric Pupil Fourier Wavefront Sensor \citep{martinache2013asymmetric} breaks the sign ambiguity in a similar way using odd pupil amplitude. In \autoref{fig:theory_fig_C} c the response of an asymmetric aperture to sign changes in odd phase aberrations is shown. It shows that, as expected, the ability of measuring odd phase modes is not affected. \autoref{fig:theory_fig_C} d has an even phase aberration, again the top and bottom rows differ in the sign. In contrast to \autoref{fig:theory_fig_C} b, the PSF for an asymmetric aperture does show a morphological change when the sign of the even aberration changes and therefore allows for the retrieval of the complete pupil-plane phase. \\ 
\\
\indent Now we will show that wavefront sensing with APPs will give similar results. In \autoref{fig:theory_fig_E} a and b, the focal-plane intensity responses are shown to an odd and even pupil-plane phase aberration for an APP designed for a symmetric aperture. Similar to \autoref{fig:theory_fig_C}, the APP shows an intensity morphology change for a sign change in the odd phase aberration (\autoref{fig:theory_fig_E} a), but for a sign change in the even phase aberration there is no intensity morphology change (\autoref{fig:theory_fig_E} b). \\
\indent This changes for an APP designed for an asymmetric aperture, as shown in \autoref{fig:theory_fig_E} c and d. \autoref{fig:theory_fig_E} d shows the focal-plane intensity response to an odd pupil-plane phase aberration and \autoref{fig:theory_fig_E} d the response to an even pupil phase aberration. Now, for both phase aberrations there is an intensity morphology change when the sign of the aberration changes.

\section{Implications for spatial LDFC}\label{sec:implications_LDFC}
Above, we discussed the principle behind FPWFS in the context of the coronagraphic PSFs of the APP, but the same principle applies to maintaining the contrast in the dark hole, such as electric field conjugation \citep{groff2015methods}, speckle nulling \citep{borde2006high} or spatial Linear Dark Field Control (LDFC; \citealt{miller2017spatial}). Spatial LDFC maintains the contrast in the dark hole by monitoring the intensity of the bright field, which can be shown to have a linear response to small phase aberrations. Suppose the focal-plane electric field ${E}_{\text{foc}}$ consists of the nominal electric field in the bright field ${E}_0$ disturbed by the electric field of the aberration ${E}_{\text{ab}}$:
\begin{align}
{E}_{\text{foc}} &= {E}_{0} + {E}_{\text{ab}}, \\
{I}_{\text{foc}} &= |{E}_{\text{foc}}|^2 \\ 
  			    &= |{E}_{0}|^2 + |{E}_{\text{ab}}|^2 + 2 \Re \{ {E}_0 {E}_{\text{ab}}^* \}. 
\end{align}
Assuming that $ |{E}_{0}|^2\gg|{E}_{\text{ab}}|^2$, we can write the intensity change $\Delta {I}$, compared to the reference image ${I}_0 = |{E}_{0}|^2$, due to the aberration as:
\begin{align}
\Delta {I} &= {I}_{\text{foc}} -  {I}_0 \\
		    &= 2 \Re \{ {E}_0 {E}_{\text{ab}}^* \} 
\end{align}
Writing the electric fields as their real and imaginary components, ${E}_0 = {a} + i {b}$ and ${E}_{\text{ab}} = {c} + i {d}$, shows that in order to have a response to the complete electric field of the aberration, the bright field of the PSF should have real and imaginary components: 
\begin{equation}
\Delta {I} = 2 ({a} {c} + {b} {d}).
\end{equation}
As discussed above, the real component ${a}$ can be provided by either a known, even phase aberration such as defocus, or an amplitude asymmetry. \\
\indent In \cite{miller2018development} spatial LDFC was tested, both in simulation and in the lab, with a vAPP designed for an even aperture ($a=0$). It was indeed observed that the LDFC loop was more stable when the vAPP image was defocussed ($a\neq0$), compared to a focussed image ($a=0$). Therefore, adding a pupil amplitude asymmetry, generating a non-zero $a$, would allow the LDFC loop to run with a focussed vAPP image, having a comparable stability as the defocussed one. 

\section{Derivatives objective function}\label{sec:derivatives}
The phase estimation is performed by minimizing the objective function $\mathcal{L}$ defined in \autoref{eq:objective_function}. The convergence speed and accuracy of the minimization algorithm are improved when it is provided with the gradients of the parameters it is estimating. For the estimation, the phase is expanded on a truncated mode basis $\{{\phi}_i \}$ with $\alpha_i$ the estimated coefficients; thus the gradient $\partial \mathcal{L} / \partial \alpha_i$ needs to be derived. Besides the phase estimation, the algorithm can also estimate the photon number $N_p$, the background level $N_b$, the fractional degree of circular polarization $v$ and the amount of leakage $L$. Therefore, the gradients $\partial \mathcal{L} / \partial N_p$, $\partial \mathcal{L} / \partial N_b$, $\partial \mathcal{L} / \partial v$ and $\partial \mathcal{L} / \partial L$ also need to be derived. \\
\indent The gradient of $\mathcal{L}$ to $X$ ($X=\{\alpha_i, N_p, N_b, v, L \}$) is given by:
\begin{equation}\label{eq:first_derivative}
\frac{\partial \mathcal{L}}{\partial X}  = \sum_{{x}} \frac{1}{\sigma_n^2} [{D} - {M}] \frac{\partial {M}}{\partial X} + \frac{\partial \mathcal{R}(\alpha)}{\partial X}.
\end{equation}
with ${M}$ the model of the system given by \autoref{eq:PSF_model}. Here, the dependency of $\mathcal{L}$ and ${M}$ on $({\alpha}, N_p, N_b, v, L)$ is omitted for readability. The term $\mathcal{R}(\alpha) / \partial X$ is only non-zero for $X=\alpha_i$.\\
\indent We start with the gradient $\partial \mathcal{L} / \partial \alpha_i$; with the first step already shown in \autoref{eq:first_derivative}, the derivative of ${M}$ is: 
\begin{equation}\label{eq:model_derivative}
\frac{\partial {M}}{\partial \alpha_i} =   N_p  \sum_{j=1}^3  a_j (v, L)  \frac{\partial {I}_{\text{foc}, j}}{\partial \alpha_i} 
\end{equation}
where the sum is over the two coronagraphic PSFs and the non-coronagraphic leakage PSF,  and ${I}_{\text{foc}, j}$ is given in \autoref{eq:focal_intensity}. The derivative $\partial {I}_{\text{foc},j} ({\alpha}) / \partial \alpha_i$ is:
\begin{equation}
\frac{\partial {I}_{\text{foc},j} ({\alpha})}{\partial \alpha_i} = \frac{\partial {E}_{\text{foc}, j}}{\partial \alpha_i} {E}_{\text{foc},j}^*+ {E}_{\text{foc},j} \frac{\partial {E}_{\text{foc},j}^*}{\partial \alpha_i},
\end{equation}
The partial derivatives to the focal-plane electric field ${E}_{\text{foc}}$ are:
\begin{align}
\frac{\partial {E}_{\text{foc}, j}}{\partial \alpha_i} &=  \frac{\partial}{\partial \alpha_i} \mathcal{C} \{ {E}_{\text{pup,j}}({\alpha}) \} \nonumber \\
							    &= \frac{\partial}{\partial \alpha_i} \mathcal{C} \{{A} e^{i ({\theta}_j + \sum_k \alpha_k {\phi}_k)} \}  \nonumber \\
							   &= \mathcal{C} \{ {A} \frac{\partial}{\partial \alpha_i}  e^{i ({\theta}_j + \sum_k \alpha_k {\phi}_k)} \}, \ \text{(Leibniz's integration rule)} \nonumber\\
							   &= i \mathcal{C} \{ {A} {\phi}_i e^{i ({\theta}_j + \sum_k \alpha_k {\phi}_k)} \}
\end{align}
The partial derivatives to the focal-plane electric field ${E}_{\text{foc}}^*$ is then simply:
\begin{equation}
\frac{\partial {E}_{\text{foc},j}^*}{\partial \alpha_i} = -i \mathcal{C} \{ {A} {\phi}_i e^{i ({\theta}_j + \sum_k \alpha_k {\phi}_k)} \}^*
\end{equation}
Combining these results, we find for $\partial {I}_{\text{foc},j} ({\alpha}) / \partial \alpha_i$: 
\begin{align}\label{eq:derivative_focal_I}
\frac{\partial {I}_{\text{foc}, j} ({\alpha})}{\partial \alpha_i} = &i \mathcal{C} \{ {A} {\phi}_i e^{i ({\theta}_j + \sum_k \alpha_k {\phi}_k)} \} \mathcal{C} \{{A} e^{i ({\theta}_j + \sum_k \alpha_k {\phi}_k)} \} ^* \nonumber \\
										     &-i \mathcal{C} \{{A} e^{i ({\theta}_j + \sum_k \alpha_k {\phi}_k)} \} \mathcal{C} \{ {A} {\phi}_i e^{i ({\theta}_j + \sum_k \alpha_k {\phi}_k)} \}^*
\end{align}
The regularization term $\mathcal{R} ({\alpha})$ that we adopted is fairly simple: 
\begin{equation}
\mathcal{R} ({\alpha}) = \frac{1}{2} \sum_{k=1}^N \frac{ \alpha_k^2 } {k^{\gamma}}, 
\end{equation}
with $\gamma$ a power that matches the known or assumed power-law distribution of the aberrations in the system. The derivative with respect to $\alpha_i$ is given by: 
\begin{equation}\label{eq:derivative_regularization}
\frac{\partial \mathcal{R} ({\alpha})}{\partial \alpha_i} = \frac{ \alpha_i} {i^{\gamma}}
\end{equation}
Combining the results in \autoref{eq:first_derivative}, \autoref{eq:model_derivative}, \autoref{eq:derivative_focal_I} and \autoref{eq:derivative_regularization} gives the final expression for $\partial \mathcal{L} / \partial \alpha_i$. \\ 
\indent Next, the derivatives of ${M}$ to $N_p$ and $N_b$ are: 
\begin{align}
\frac{\partial {M}}{\partial N_p} &= \sum_{j=1}^3 a_j (v, L) {I}_{\text{foc}, j} ({\alpha}), \\ 
\frac{\partial {M}}{\partial N_b} &= 1
\end{align}
The derivative of ${M}$ to the leakage $L$ is: 
\begin{equation}
\frac{\partial {M}}{\partial L} = N_p \sum_{j=1}^3  \frac{\partial a_j (v, L)}{\partial L} {I}_{\text{foc}, j},
\end{equation}
with the derivatives $\partial a_j (v, L) / \partial L$ given by:
\begin{align}
\frac{\partial a_1}{\partial L} &= -\frac{1+v}{2}, \\
\frac{\partial a_2}{\partial L} &= \frac{v-1}{2}, \\
\frac{\partial a_3}{\partial L} &= 1.
\end{align}
Finally, the derivative of ${M}$ to $v$ is: 
\begin{equation}
\frac{\partial {M}}{\partial v} =  N_p \sum_{j=1}^3  \frac{\partial a_j (v, L)}{\partial v} {I}_{\text{foc}, j},
\end{equation}
with the derivatives $\partial a_j (v, L) / \partial v$ given by:
\begin{align}
\frac{\partial a_1}{\partial v} &= \frac{1-L}{2} \\
\frac{\partial a_2}{\partial v} &= \frac{L-1}{2} \\
\frac{\partial a_3}{\partial v} &= 0
\end{align}

\end{appendix}

\end{document}